\documentclass[11pt]{article}
\usepackage{latexsym} 
\usepackage{amsfonts}
\usepackage{amsmath}
\usepackage{amsthm}
\usepackage{amssymb}

\def\subeqnarray{\arraycolsep1pt
    \def\@eqnnum\stepcounter##1{\stepcounter{subequation}%
        {\reset@font\rm(\theequation\alph{subequation})}}
\jot5mm     \eqnarray}

\makeatother

\def\be{\begin{equation}}
\def\lb#1{\label{#1}}
\def\ee{\end{equation}}
\def\bea{\begin{eqnarray}}
\def\eea{\end{eqnarray}}
\def\ba{\begin{array}}
\def\ea{\end{array}}
\def\dd{\partial}

\def\one#1{#1^{\raise5pt\hbox{$\scriptstyle\!\!\!\!1$}}\,{}}
\def\two#1{#1^{\raise5pt\hbox{$\scriptstyle\!\!\!\!2$}}\,{}}

\def\tilde{\widetilde}
\def\II{\hbox{{1}\kern-.25em\hbox{l}}}
\def\p#1{(\ref{#1})}
\makeatletter
\def\binrel@#1{\begingroup
  \setboxz@h{\thinmuskip0mu
    \medmuskip\m@ne mu\thickmuskip\@ne mu
    \setbox\tw@\hbox{$#1\m@th$}\kern-\wd\tw@
    ${}#1{}\m@th$}%
  \edef\@tempa{\endgroup\let\noexpand\binrel@@
    \ifdim\wdz@<\z@ \mathbin
    \else\ifdim\wdz@>\z@ \mathrel
    \else \relax\fi\fi}%
  \@tempa
}
\let\binrel@@\relax
\def\overset#1#2{\binrel@{#2}%
  \binrel@@{\mathop{\kern\z@#2}\limits^{#1}}}
\def\underset#1#2{\binrel@{#2}%
  \binrel@@{\mathop{\kern\z@#2}\limits_{#1}}}
\makeatother
\newfont{\bbd}{msbm10 scaled\magstep1}
\def\C{\hbox{\bbd C}}

\def\R{\hbox{\bbd R}}

\def\V{\hbox{\bbd V}}

\def\T{\hbox{\bbd T}}

\def\Z{\hbox{\bbd Z}}

\def\mR{\mathrm{R}}
\def\mL{\mathrm{L}}

\begin{document}


\begin{center}
{\LARGE \bf{New elliptic solutions of the Yang-Baxter equation}}

\vspace{1cm}

{\large \sf D. Chicherin$^{a}$\footnote{\sc e-mail:chicherin@lapth.cnrs.fr},
  S. E. Derkachov$^{b}$\footnote{\sc e-mail:derkach@pdmi.ras.ru},
 V. P. Spiridonov$^c$\footnote{\sc e-mail:spiridon@theor.jinr.ru} \\
}

\vspace{0.5cm}

\begin{itemize}
\item[$^a$]
{\it LAPTH\,\footnote{UMR 5108 du CNRS, associ\'ee \`a l'Universit\'e de Savoie}, Universit\'{e} de Savoie, CNRS,
B.P. 110,  F-74941 Annecy-le-Vieux, France}
\item[$^b$]
{\it St. Petersburg Department of Steklov Mathematical Institute
of Russian Academy of Sciences,
Fontanka 27, 191023 St. Petersburg, Russia}
\item[$^c$]
{\it Laboratory of Theoretical Physics, JINR, Dubna, Moscow region, 141980, Russia}
\end{itemize}
\end{center}
\vspace{0.5cm}
\begin{abstract}
We consider finite-dimensional reductions of an integral operator with the elliptic
hypergeometric kernel describing the most general
known solution of the Yang-Baxter equation with a rank 1 symmetry algebra.
The reduced R-operators reproduce at their bottom the
standard Baxter's R-matrix for the 8-vertex model and
Sklyanin's L-operator. The general formula has a remarkably
compact form and yields new elliptic solutions of the Yang-Baxter
equation based on the finite-dimensional representations of
the elliptic modular double. The same result is also derived using
the fusion formalism.

\end{abstract}



{\small \tableofcontents}
\renewcommand{\refname}{References.}
\renewcommand{\thefootnote}{\arabic{footnote}}
\setcounter{footnote}{0} \setcounter{equation}{0}

\section{Introduction}

Exactly solvable models play an important role in the investigation
of critical phenomena in statistical mechanics \cite{Baxter}.
One of the key structural elements leading to exact solvability is
the Yang-Baxter equation (YBE) \cite{Jimbo,TF}. In a sense, YBE
reduces many-body problems to considerations of pairwise two-body interacting systems.
In the context of discrete spin lattice systems, solutions of the YBE
are given by ordinary finite rank matrices with matrix elements
expressed either in terms of the rational, trigonometric (or hyperbolic)
or elliptic functions. In the context of quantum field theories (the continuous
spin systems), the YBE is solved in terms of the integral operators
associated with the plain hypergeometric functions, their various
$q$-analogues (trigonometric or hyperbolic) or the elliptic hypergeometric
integrals \cite{spi:essays}.

The first elliptic solution of YBE given by a $4\times 4$
matrix was derived by Baxter \cite{Baxter1}. It played a crucial role in solving
the 8-vertex model, which is related also to the XYZ spin chain \cite{Baxter,TF}.
A different $4\times 4$ R-matrix with elliptic entries was found
by Felderhof \cite{Fel}.
As shown by Krichever \cite{Kri}, the
Baxter and Felderhof R-matrices exhaust all $4 \times 4$ matrix solutions of YBE.
Elliptic R-matrices with higher rank symmetry algebras were
constructed by Belavin \cite{Belavin}.

The concept of the quantum group $U_q(s\ell_2)$ naturally arises
from a degeneration of Baxter's R-matrix to the trigonometric level \cite{Jimbo}.
However, as shown by Faddeev \cite{fad:mod}, in general it is necessary to
deal with a more complicated algebra called the modular double.
At the elliptic level corresponding algebraic structures are described
by the  Sklyanin algebra \cite{skl1,skl2} and the elliptic modular double \cite{AA2008}.
Various connections between the Sklyanin algebra and elliptic hypergeometric
functions were considered in \cite{Rains,ros:elementary,ros:sklyanin,AA2008}.

One can construct higher dimensional matrix solutions of YBE out of the fundamental
ones using the fusion procedure  \cite{KRS81}, which was applied to many
situations. For instance, fusion of elliptic R-matrices for the SOS type
models has been constructed by the Kyoto group \cite{DJMO86,DJKMO87,DJKMO88}.
This procedure yielded rather complicated forms of the Boltzmann weights which
appeared to be combinable to a nice elliptic hypergeometric series \cite{FT}.
The fusion was applied also to the Sklyanin algebra
representations, see e.g. \cite{McCFa04,Hasegawa97,HZ90,konno,konno2,tak,Tak96}
and references therein.

Integrable systems with the continuous spin variables (noncompact spin chains)
are important in quantum field theory. A systematic approach to solving
YBEs for such systems based on a twisted representation of the permutation group
generators by integral operators was developed in some of our previous papers.
Construction of an elliptic hypergeometric solution of the YBE following this line
and using some formal infinite series was given in  \cite{Derkachov:2007gr}.
A similar construction based on the technique of
intertwining vectors was described later in \cite{Zab11}.
The most complicated known solution of YBE (with the rank 1 symmetry algebra)
as an integral operator with the elliptic hypergeometric kernel
has been constructed in \cite{DS} following the logical line of
\cite{Derkachov:2007gr} boosted by a powerful machinery of
elliptic hypergeometric integrals \cite{spi:umn,AA2003,spi:essays}.
The key ingredients in this construction are
the elliptic beta integral evaluation formula discovered in \cite{spi:umn}
and the integral operator introduced in \cite{spi:bailey},
which appeared to be the intertwining operator for the Sklyanin algebra.
The integral Bailey lemma proven in \cite{spi:bailey}
yields the star-triangle relation in the operator form.
Some further developments of the results of \cite{DS} are described in \cite{CDKKIII,DS2}.
In particular, in  \cite{DS2} finite-dimensional representations of
the elliptic modular double have been considered in a systematic way.

Various finite-dimensional R-matrices and L-operators
were obtained in \cite{CDS1} from reductions of general integral operator solutions
of the YBE. In the rational case this procedure yields  perhaps all
such finite-dimensional R-matrices. In the $q$-deformed cases
the situation is more complicated. Namely, such
finite-dimensional solutions of YBE were built for the plain $U_q(s\ell_2)$-algebra
and the Faddeev modular double cases under the restriction
that $q$ is not a root of unity.
The principal aim of the present paper is the construction of new
elliptic solutions of the YBE.  They are characterized
by the presence of some two-dimensional discrete lattices for each spin variable leading
to ``a doubling" of the dimension of the finite-dimensional representation space.
Existence of such ``two-index" solutions of the YBE was conjectured
in \cite{spi:kiev} on the basis of some properties of the
elliptic hypergeometric integrals leading to a new class of biorthogonal functions
of hypergeometric type \cite{AA2003,spi:essays}.
First, we derive these solutions with a slightly formal approach as reductions
of the general integral R-operator. Then we confirm this result in a rigorous way
by a completely different procedure called the fusion. In the latter
considerations, our approach is most close
to the ones developed by Takebe \cite{tak,Tak96} and Konno \cite{konno2}
(which is based on Rosengren's results \cite{ros:elementary}).

The plan of the paper is as follows. We start with a short review of the well-know facts
about elliptic solutions of YBE. In Sect. 2 we present Baxter's R-matrix for the 8-vertex
model and describe some basic properties of the Jacobi theta functions.
In Sect. 3 we review the Sklyanin algebra, its finite-dimensional and infinite-dimensional
representations, and corresponding Lax operator.
In particular, we show that the reduction of the Lax operator to two-dimensional representation
in the quantum space
coincides with Baxter's R-matrix. In Sect. 4 we proceed to the elliptic modular double.
Our considerations are heavily based on the intertwining operator of equivalent representations
of this algebra. It serves as a basic tool enabling us
to describe finite-dimensional representations.
Moreover, it suggests a natural way for constructing the generating function of
finite-dimensional representations embracing all basis vectors of
the representation space.
In the beginning of Sect. 5 we briefly outline derivation of the general R-operator which
acts in the tensor product of two infinite-dimensional representations of the elliptic modular double.
Then, in Sect. 5.1, we proceed to the main topic of the paper. There we calculate reductions
of the general R-operator to finite-dimensional matrices in one of the spaces
as well as in both spaces. The results are described by the remarkably
compact formulae \p{redsl2''} and \p{redsl2}.
They comprise an enormous number of elliptic YBE solutions, both plain R-matrices
and L-operators with the elliptic function entries.
In order to demonstrate the power of the reduction formulae
we recover Sklyanin's L-operator out of the general R-operator in Sect. 5.2.
Then, in Sect. 6, we demonstrate how to reproduce these results by means of the fusion method.
In Sect. 6.1 we fuse an arbitrary number of Baxter's R-matrices
and explicitly reconstruct Sklyanin's L-operator for finite-dimensional representations.
In Sect. 6.2 we proceed to a higher level of complexity.
There we accomplish the fusion of arbitrary number of Lax operators for
infinite-dimensional representations.
The result is a higher-spin R-operator -- a generalization of Sklyanin's L-operator
-- which is defined on the tensor product of an infinite-dimensional
representation and arbitrary finite-dimensional representation.
The derived formula is in a nice agreement with the reduction result from Sect. 5.1.
We conclude in Sect. 7 where we discuss possible applications of the obtained solutions of YBE.

\section{Baxter's R-matrix}
\label{Baxter} \setcounter{equation}{0}

The Yang-Baxter equation with the spectral parameter which we investigate
in this work has the form
\begin{equation}\label{YB}
\mathbb{R}_{12} (u-v)\,\mathbb{R}_{13}(u)\, \mathbb{R}_{23}(v)
=\mathbb{R}_{23}(v)\,\mathbb{R}_{13}(u)\,\mathbb{R}_{12}(u-v)\, .
\end{equation}
Its solutions are called the R-matrices or R-operators. The operators $\mathbb{R}_{ik}(u)$
depend on the complex spectral parameter $u\in \mathbb{C}$ and act in the
tensor product of three (in general different) spaces $\V_1\otimes\V_2\otimes\V_3$.
The indices $i$ and $k$ indicate that $\mathbb{R}_{ik}(u)$ acts nontrivially in the
subspace $\V_i\otimes\V_k$ and it is the unity operator in the remaining
part of $\V_1\otimes\V_2\otimes\V_3$. In the most general situation all
three spaces $\V_i$ are infinite-dimensional and $\mathbb{R}_{ik}(u)$
are described by integral operators with some singular kernels.

The equation \p{YB} is inherently universal since it does not involve
information neither on the symmetry algebra underlying an integrable model
nor on its particular representations in the spaces $\V_i$.
In this paper we deal with solutions of \p{YB}
associated with an elliptic deformation of a rank 1 Lie algebra to be specified below.
Further we specify spaces $\V_i$ in \p{YB} and corresponding representations
of the symmetry algebra starting from two-dimensional representations and
heading towards infinite-dimensional ones. In this way we go through the hierarchy of
elliptic solutions of YBE from the simplest to the most intricate R-matrices.

In the simplest case
all $\V_i$-spaces in \p{YB} are two-dimensional, $\V_i =\mathbb{C}^2$\,.
A corresponding elliptic R-matrix has been found by Baxter in solving
the eight-vertex model~\cite{Baxter1,Baxter,FT}
\begin{equation}\label{RBaxter}
\mathbb{R}_{12}(u) = \sum_{\alpha=0}^3 w_{\alpha} (u)\,
\sigma_{\alpha} \otimes\sigma_{\alpha} = \left(
\begin{array}{cc}
w_0(u)\,\sigma_0+w_3(u)\,\sigma_3 &
w_1(u)\,\sigma_1-\textup{i} w_2(u)\,\sigma_2 \\
w_1(u)\,\sigma_1+\textup{i} w_2(u)\,\sigma_2&
w_0(u)\,\sigma_0-w_3(u)\,\sigma_3
\end{array} \right)\,,
\end{equation}
where $\sigma_0=1$ and $\sigma_{\alpha}\,,\, \alpha=1,2,3,$ are the standard Pauli matrices and
the coefficient functions depend on the spectral parameter
$$
w_{\alpha}(u) = \frac{\theta_{\alpha+1}(u+\eta)}{\theta_{\alpha+1}(\eta)}.
$$
We use the shorthand notation $\theta_{\alpha}(u)\equiv \theta_{\alpha}(u|\tau)$ for
Jacobi theta-functions with the modular parameter $\tau$
\begin{eqnarray}\label{theta1} &&
\theta_{1}(z|\tau) = -\sum_{n\in\mathbb{Z}}
\mathrm{e}^{\pi \textup{i} \left(n+\frac{1}{2}\right)^2\tau}\cdot
\mathrm{e}^{2\pi \textup{i}
\left(n+\frac{1}{2}\right)\left(z+\frac{1}{2}\right)}
=\frac{e^{-\pi\textup{i}z} \theta(e^{2\pi \textup{i}z};p)}{\mathrm{R}(\tau)}, \quad
\mathrm{R}(\tau) = \frac{p^{-\frac{1}{8}}}
{\textup{i} (p;p)_\infty},
\end{eqnarray}
where in the multiplicative notation
\be \lb{multnot}
p=e^{2\pi \textup{i} \tau} \;\;,\;\;
\theta(t;p)=(t;p)_\infty(pt^{-1};p)_\infty \;\;,\;\;
(t;p)_\infty=\prod_{k=0}^\infty(1-tp^k)\,.
\ee
The other three theta-functions are obtained by shifts of the argument of
$\theta_1$ by quasiperiod halves
\begin{eqnarray*} &&
\theta_{2}(z|\tau)=\theta_1(z+{\textstyle\frac{1}{2}}|\tau)\,, \quad
\theta_{3}(z|\tau)=e^{\frac{\pi \textup{i}\tau}{4}+\pi \textup{i} z}\theta_2(z+{\textstyle \frac{\tau}{2}}|\tau)\,, \quad
\theta_4(z|\tau)= \theta_3(z+{\textstyle\frac{1}{2}}|\tau)\,.
\end{eqnarray*}
Quasiperiodic properties of these functions are described by the relations
$$
\theta_{1}(z+1|\tau) =-\theta_{1}(z|\tau)\,, \quad
\theta_{1}(z+\tau|\tau) =- e^{-2\pi\textup{i} z-\pi\textup{i}\tau}\theta_{1}(z|\tau)\,.
$$
Other indispensable elliptic special functions and some
necessary identities for theta-functions will be indicated in proper places below.

The Baxter R-matrix \p{RBaxter} depends on the spectral parameter $u\in\mathbb{C}$
and two additional free variables $\eta, \,\tau\in\mathbb{C}$, such that
$\theta_\alpha(\eta)\neq 0,\, \alpha=1,\ldots,4,$ and Im$(\tau)>0\,$. Its connection
to a symmetry algebra and representation theory is explained in the next section.

The R-matrix \p{RBaxter} is not the only $4 \times 4$ matrix solution of the YBE.
As mentioned already, there exists a different solution of YBE for
$\V_i =\mathbb{C}^2$, which has been found by Felderhof~\cite{Fel}.
In this paper we concentrate on the solutions of YBE associated with
Baxter's case and postpone the study of Felderhof's solution.

\section{L-operator and the Sklyanin algebra}
\label{Sklyanin}

\subsection{Algebraic relations}

At the next level of complexity of YBE
\eqref{YB} one of the spaces, say $\V_3$, is arbitrary, and
the remaining two spaces are 2-dimensional $\V_1 = \mathbb{C}^2$, $\V_2 =\mathbb{C}^2$. In this case the R-matrix
$\mathbb{R}_{13}(u)\equiv\mathrm{L}_{13}(u)$ (and $\mathbb{R}_{23}(u)
\equiv\mathrm{L}_{23}(u)$)
is known as the quantum L-operator or the Lax matrix.
It acts as a $2 \times 2$ matrix constructed out of the Pauli matrices
in the space $\V_1$,
\begin{equation}\makebox[-1em]{}
\mathrm{L}_{13}(u)=\mathrm{L}(u) \equiv
\sum_{\alpha=0}^3 w_{\alpha} (u)\, \sigma_\alpha \otimes
\mathbf{S}^{\alpha} = \left(
\begin{array}{cc}
w_0(u)\,\mathbf{S}^0+w_3(u)\,\mathbf{S}^3 &
w_1(u)\,\mathbf{S}^1-\textup{i} w_2(u)\,\mathbf{S}^2 \\
w_1(u)\,\mathbf{S}^1+\textup{i} w_2(u)\,\mathbf{S}^2&
w_0(u)\,\mathbf{S}^0-w_3(u)\,\mathbf{S}^3
\end{array} \right),
\label{L_op}\end{equation}
whose $u$-independent entries $\mathbf{S}^{\alpha}$ are some operators acting in $\V_3$.
The functions $w_{\alpha}(u)$ are the same as in the Baxter R-matrix \p{RBaxter}.
The space $\V_1 = \mathbb{C}^2$ in \p{L_op} is usually referred to as an auxiliary space, and
$\V_3$ is called the quantum space of the Lax operator.
The same operators $\mathbf{S}^{\alpha}$
enter the second copy of the Lax operator $\mathrm{L}_{23}(u)$,
which acts as a $2\times 2$ matrix in the space  $\V_2 = \mathbb{C}^2$.
In this setting equation \p{YB} takes the form of a RLL-relation \cite{TF}
\begin{equation}\label{Lax}
\mathbb{R}_{12} (u-v)\,\mathrm{L}_{13}(u)\, \mathrm{L}_{23}(v)
=\mathrm{L}_{23}(v)\,\mathrm{L}_{13}(u)\,\mathbb{R}_{12}(u-v)\,,
\end{equation}
where $\mathbb{R}_{12}(u)$ is Baxter's R-matrix~(\ref{RBaxter}).
Since the operator $\mathbb{R}_{12}$ in \p{Lax} has been already specified, this
relation can be considered as a nontrivial restriction for the operators
$\mathbf{S}^{\alpha}$ and the space $\V_3$ where the operators $\mathbf{S}^{\alpha}$
are acting. After some laborious work one can see that equation
\eqref{Lax} is equivalent to the following set of commutation relations for
four operators $\mathbf{S}^0, \mathbf{S}^1,\mathbf{S}^2,\mathbf{S}^3$
forming the Sklyanin algebra \cite{skl1,skl2}:
\begin{align}
\mathbf{S}^\alpha\,\mathbf{S}^\beta - \mathbf{S}^\beta\,\mathbf{S}^\alpha =
\textup{i}\left(\mathbf{S}^0\,\mathbf{S}^\gamma +\mathbf{S}^\gamma\,\mathbf{S}^0\right)\,,
\notag \\
\mathbf{S}^0\,\mathbf{S}^\alpha - \mathbf{S}^\alpha\,\mathbf{S}^0 =
\textup{i}\,\mathbf{J}_{\beta \gamma}\bigl(\mathbf{S}^\beta\,\mathbf{S}^\gamma +\mathbf{S}^\gamma\,\mathbf{S}^\beta \bigr)\,,
\lb{SklAlg}
\end{align}
where the triplet $(\alpha,\beta,\gamma)$ is an arbitrary cyclic permutation
of $(1,2,3)$. The structure constants $\mathbf{J}_{\beta \gamma}$ are
parametrized in terms of theta functions as
\begin{equation}\label{Jik}
\mathbf{J}_{12}=\frac{\theta_1^2(\eta)\theta_4^2(\eta)}
{\theta_2^2(\eta)\theta_3^2(\eta)},\qquad
\mathbf{J}_{23}=\frac{\theta_1^2(\eta)\theta_2^2(\eta)}
{\theta_3^2(\eta)\theta_4^2(\eta)}, \qquad
\mathbf{J}_{31}= -\frac{\theta_1^2(\eta)\theta_3^2(\eta)}
{\theta_2^2(\eta)\theta_4^2(\eta)}\,
\end{equation}
and satisfy the constraint $\mathbf{J}_{12}+\mathbf{J}_{23}+\mathbf{J}_{31}
+\mathbf{J}_{12}\mathbf{J}_{23}\mathbf{J}_{31}=0$.
One can write $\mathbf{J}_{\alpha\beta}=
\frac{\mathbf{J}_{\beta}-\mathbf{J}_{\alpha}}{\mathbf{J}_{\gamma}}$,
$\gamma\neq \alpha,\beta$, where
\be \lb{strcnst}
\mathbf{J}_{1}=\frac{\theta_2(2\eta)\theta_2(0)}
{\theta_2^2(\eta)},\qquad
\mathbf{J}_{2}=\frac{\theta_3(2\eta)\theta_3(0)}
{\theta_3^2(\eta)}, \qquad
\mathbf{J}_{3}= \frac{\theta_4(2\eta)\theta_4(0)}
{\theta_4^2(\eta)}\,.
\ee

Casimir operators are indispensable for classification of irreducible representations of any algebra.
The Sklyanin algebra has two Casimir operators commuting with all generators:
$\left[\mathbf{K}_0 ,\mathbf{S}^{\alpha}\right] =
\left[\mathbf{K}_2 ,\mathbf{S}^{\alpha}\right] =0$\,,
\be \lb{Cas1}
\mathbf{K}_0 = \sum_{\alpha=0}^3\,\mathbf{S}^{\alpha}\,\mathbf{S}^{\alpha},\qquad
\mathbf{K}_2 = \sum_{\alpha=1}^3\,\mathbf{J}_{\alpha}\,\mathbf{S}^{\alpha}\,\mathbf{S}^{\alpha}\,.
\ee
A remarkable feature of the algebraic structure \p{SklAlg} is that it admits a highly nontrivial
explicit realization of generators as finite-difference
operators with elliptic coefficients found by Sklyanin in his pioneering paper \cite{skl2}
\begin{equation}\label{Sklyan1}
\mathbf{S}^{\alpha} =\frac{\textup{i}^{\delta_{\alpha,2}}
\theta_{\alpha+1}(\eta)}{\theta_1(2 z) } \Bigl[\,\theta_{\alpha+1} \left(2
z-2\eta\ell\right)\,e^{\eta\partial} - \theta_{\alpha+1}
\left(-2z-2\eta\ell\right)\,e^{-\eta\partial}\, \Bigl]\,,
\end{equation}
where $\mathrm{e}^{\eta\dd}$ is a shift operator, $\mathrm{e}^{\eta\dd}f(z)=f(z+\eta).$
The test functions $f(z)$ must be in general meromorphic, since the operators $\mathbf{S}^{\alpha}$
have meromorphic coefficients.

The variable $\ell\in\mathbb{C}$ is called the {\em spin}.
In this realization the Casimir operators reduce to the following
scalar expressions
$$
\mathbf{K}_0 = 4\,\theta_1^2\bigl(2\ell\eta+\eta\bigr),\qquad
\mathbf{K}_2 = 4\,\theta_1\bigl(2\ell\eta+2\eta\bigr)\,
\theta_1\bigl(2\ell\eta\bigr)\,.
$$
The spin $\ell$ labels the Sklyanin algebra representations since it fixes
(together with $\eta$ and $\tau$) the Casimir operator values.

There exists a useful factorized representation for the Lax operator $\mathrm{L}(u)$  \p{L_op}
when the operators $\mathbf{S}^{\alpha}$ are given by the explicit expression \eqref{Sklyan1}
 \cite{Derkachov:2007gr,KrZa97,Zab99,Zab11},
\be \lb{LFact}
\mathrm{L}(u_1,u_2) = \frac{1}{\theta_1(2 z)} \left(
\begin{array}{cc}
\bar{\theta}_3\left(z - u_1\right) & -\bar{\theta}_3\left(z+u_1\right) \\
-\bar{\theta}_4\left(z - u_1\right) & \bar{\theta}_4\left(z+u_1\right)
\end{array} \right)
\left(
\begin{array}{cc}
\mathrm{e}^{\eta \dd} &0\\
0 & \mathrm{e}^{-\eta \dd}
\end{array} \right )
\left(
\begin{array}{cc}
\bar{\theta}_4\left(z+u_2\right) & \bar{\theta}_3\left(z+u_2\right) \\
\bar{\theta}_4\left(z - u_2\right) & \bar{\theta}_3\left(z - u_2\right)
\end{array} \right)\,.
\ee
Here we use the shorthand notation
$\bar\theta_{\alpha}(z)\equiv \theta_{\alpha}(z|{\textstyle\frac{\tau}{2}})$ for
theta-functions with the modular parameter ${\textstyle\frac{\tau}{2}}$.
Instead of working with the spectral parameter $u$ and the spin $\ell$ separately
we prefer to combine them to a pair of new ``light-cone" parameters
which are linear combinations of the latter
\be \lb{u1u2}
u_1 = \frac{u}{2} + \eta\ell + \frac{\eta}{2}, \qquad
u_2 = \frac{u}{2} - \eta\ell - \frac{\eta}{2}\,.
\ee
The factorization formula for the Lax operator plays an essential role in
the construction of the general $\mathrm{R}$-operator
(see \cite{DS} and Sect.~\ref{SecR}) and it is crucial in the ``fusion"
of Baxter's R-matrices (see Sect.~\ref{SecFus}).

\subsection{Finite-dimensional representations of the Sklyanin algebra}

For generic values of the spin $\ell$ the Sklyanin algebra representation
determined by the generators \p{Sklyan1} is irreducible and infinite-dimensional.
It can be realized in the space of meromorphic functions of one complex variable $z$.
However, for a discrete set of spin values, say, $2\ell=n$, $n\in\Z_{\geq 0}$,
finite-dimensional representations arise
and the action of generators \p{Sklyan1} leaves invariant the corresponding
finite-dimensional subspace of holomorphic functions.
Indeed, for the (half)-integer spin $\ell =\frac{n}{2}$ the irreducible representation
is $(n+1)$-dimensional and it can be realized in the space $\Theta^{+}_{2n}$
consisting of even theta functions of order $2n$ \cite{skl2}.
The space $\Theta^+_{2n}$ is formed by the holomorphic functions
of complex variable $z$ which are even $f(z)=f(-z)$ and have
simple quasiperiodicity properties under the shifts of $z$ by $1$ and $\tau$:
$$
f(z+1) = f(z), \qquad f(z+\tau) =
\mathrm{e}^{-2 n\pi i\tau -4 n\pi i z } f(z)\,.
$$

Further, it will be useful for us to choose a particular basis in the space $\Theta^{+}_{2n}$.
Let us consider monomials in a pair of theta functions $\bar\theta_4(z)$, $\bar\theta_3(z)$
with the quasiperiods $1$ and $\tau/2$.
Taking into account their quasiperiodicity properties,
we see that the following set of $n+1$ functions belongs to the space $\Theta^+_{2n}$
and, due to the linear independence, forms a basis
\be \lb{basisSkl}
\bar\theta_3^j \left(z\right) \,
\bar\theta_4^{n-1-j} \left(z\right), \quad j=0,1,\cdots, n\,.
\ee

In order to clarify how Baxter's R-matrix \p{RBaxter} is related to the Sklyanin algebra
representations let us consider in more detail the two-dimensional case corresponding
to the spin $\ell =1/2$.
In that case the Sklyanin algebra generators \p{Sklyan1} take the following form
\begin{equation} \lb{Skl12}
\left[\,\mathbf{S}^{\alpha}\,\Phi\,\right](z) =\frac{(\textup{i})^{\delta_{\alpha,2}}
\theta_{\alpha+1}(\eta)}{\theta_1(2 z) } \Bigl[\,\theta_{\alpha+1} \left(2
z-\eta\right)\cdot \Phi(z+\eta) - \theta_{\alpha+1}
\left(-2z-\eta\right)\cdot \Phi(z-\eta)\, \Bigl]\,.
\end{equation}
Let us restrict them to the two-dimensional space $\Theta^+_2$
using the basis $\mathbf{e}_1=\bar\theta_4(z)$ and $\mathbf{e}_2=\bar\theta_3(z)$ \p{basisSkl}.
To do this, we compute the action of generators \p{Skl12} on this basis vectors
and expand the result in the same basis. Using the following identities for theta functions
\begin{eqnarray}\nonumber &&
2\,\theta_1(x+y)\,\theta_1(x-y) = \bar\theta_4(x)\,\bar\theta_3(y)
-\bar\theta_4(y)\,\bar\theta_3(x),
\\ \nonumber &&
2\,\theta_2(x+y)\,\theta_2(x-y) = \bar\theta_3(x)\,\bar\theta_3(y)
-\bar\theta_4(y)\,\bar\theta_4(x),
\\ \label{identities} &&
  2\,\theta_3(x+y)\,\theta_3(x-y)
= \bar\theta_3(x)\,\bar\theta_3(y) +\bar\theta_4(y)\,\bar\theta_4(x),
\\ \nonumber &&
2\, \theta_4(x+y)\,\theta_4(x-y)=\bar\theta_4(x)\,\bar\theta_3(y)
+\bar\theta_4(y)\,\bar\theta_3(x),
\end{eqnarray}
one obtains in a straightforward fashion
\begin{eqnarray*}
&& \mathbf{S}^0\,\bar\theta_3(z) = \bar\theta_3(z)\, \theta_1(2\eta),\quad \quad
\mathbf{S}^0\,\bar\theta_4(z) = \bar\theta_4(z)\, \theta_1(2\eta)\,,
\\ 
&& \mathbf{S}^1\,\bar\theta_3(z) = \bar\theta_4(z)\, \theta_1(2\eta),\quad \quad
\mathbf{S}^1\,\bar\theta_4(z) = \bar\theta_3(z)\, \theta_1(2\eta)\,,
\\ 
&& \mathbf{S}^2\,\bar\theta_3(z) = -\textup{i}\,\bar\theta_4(z)\, \theta_1(2\eta),\quad
\mathbf{S}^2\,\bar\theta_4(z) = \textup{i}\,\bar\theta_3(z)\, \theta_1(2\eta)\,,
\\ 
&& \mathbf{S}^3\,\bar\theta_3(z) = -\bar\theta_3(z)\, \theta_1(2\eta),\quad \;
\mathbf{S}^3\,\bar\theta_4(z) = \bar\theta_4(z)\, \theta_1(2\eta)\,.
\end{eqnarray*}
These relations can be arranged to a matrix form.
Consequently, we find that the Sklyanin algebra generators \p{Skl12} are reduced
to the Pauli sigma-matrices in the basis \p{basisSkl},
\be \lb{Sred}
\mathbf{S}^\alpha\, \left(\mathbf{e}_1\,, \mathbf{e}_2\right) =
\left(\mathbf{S}^\alpha\,\mathbf{e}_1\,, \,\mathbf{S}^\alpha\,\mathbf{e}_2\right) =
\left(\mathbf{e}_1\,,\, \mathbf{e}_2\right)\, \sigma^{\alpha}\,\theta_1(2\eta)\,.
\ee
Finally, we apply this reduction \p{Sred} to the Lax
operator \p{L_op} action in the quantum space and find that it becomes
proportional\footnote{We warn the reader that due to an
unconventional choice of the spectral parameter shift in \p{L_op} (cf. equations
(2.5) and (2.6) in \cite{Derkachov:2007gr}) there is no discrepancy between dependences
on the spectral parameter $u$ of the Baxter R-matrix and the Lax operator reduction.}
to Baxter's R-matrix $\mathbb{R}(u)$ \p{RBaxter},
\begin{eqnarray} \nonumber &&
\mathrm{L}(u)\left(\mathbf{e}_1\,, \mathbf{e}_2\right) =
\sum_{\alpha=0}^3 w_{\alpha} (u)\, \sigma_\alpha \otimes\,
\left(\mathbf{S}^\alpha\,\mathbf{e}_1\,,\, \mathbf{S}^\alpha\,\mathbf{e}_2\right) =
\\ && \makebox[5em]{}
=\theta_1(2\eta)\, \sum_{\alpha=0}^3 w_{\alpha} (u)\, \sigma_\alpha \otimes\,
\left(\mathbf{e}_1\,,\, \mathbf{e}_2\right) \sigma^{\alpha} =
\theta_1(2\eta)\,\left(\mathbf{e}_1\,, \mathbf{e}_2\right)\,
\mathbb{R}_{12}(u)\,.\lb{Lred}
\end{eqnarray}
Thus the Baxter R-matrix is incorporated to the set of RLL-relation solutions \p{L_op}
and its entries provide two-dimensional (the fundamental) representation
of the Sklyanin algebra \p{SklAlg}. In Sect. \ref{SecFus} we follow an opposite strategy
and show explicitly how Lax operators with arbitrary finite-dimensional representations
in the quantum space can be ``fused" out of Baxter's R-matrix \p{RBaxter}.

\section{The elliptic modular double}
\lb{SecDoub}

\subsection{Algebraic relations}

In this section we consider an algebra which is called the elliptic modular double
and which, roughly speaking, consists of two Sklyanin algebras.
First, we  outline the structure of finite-dimensional representations
of this algebra postponing corresponding solutions of YBE \p{YB} to the next section.
The necessity of the doubling of Sklyanin algebras will be clarified there. We will
see that the symmetry requirements with respect to the
extended algebra pose additional restrictions, as compared to the Sklyanin algebra,
that enable us to fix uniquely the YBE solutions.

The elliptic modular double was introduced in \cite{AA2008}.
A particular version of this algebra degenerates in a special limit
to Faddeev's modular double of the quantum algebra $U_q(s\ell_2)$ \cite{fad:mod}.
Let us take two sets of generators
$\mathbf{S}^\alpha$ and  $\mathbf{\tilde S}^\alpha$ both respecting
Sklyanin algebra commutation relations.
The generators $\mathbf{S}^\alpha$ fulfill commutation relations \p{SklAlg},
and the operators $\mathbf{\tilde S}^\alpha$ satisfy analogues of \p{SklAlg}
with a different set of structure constants $\mathbf{\tilde J}_{\alpha\beta}$, i.e.
\begin{eqnarray}\nonumber &&
\mathbf{\tilde S}^\alpha\,\mathbf{\tilde S}^\beta - \mathbf{\tilde S}^\beta\,\mathbf{\tilde S}^\alpha =
\textup{i}\left(\mathbf{\tilde S}^0\,\mathbf{\tilde S}^\gamma +\mathbf{\tilde S}^\gamma\,\mathbf{\tilde S}^0\right)\,,
\\ &&
\mathbf{\tilde S}^0\,\mathbf{\tilde S}^\alpha - \mathbf{\tilde S}^\alpha\,\mathbf{\tilde S}^0 =
\textup{i}\,\mathbf{\tilde J}_{\beta \gamma}\left(\mathbf{\tilde S}^\beta\,\mathbf{\tilde S}^\gamma
+\mathbf{\tilde S}^\gamma\,\mathbf{\tilde S}^\beta\right)\,,
\label{sklalg_doub}\end{eqnarray}
where the triplet $(\alpha,\beta,\gamma)$ is an arbitrary cyclic permutation
of $(1,2,3)$. The modular double which we shall be using below corresponds to
the $\mathbf{\tilde S}^\alpha$-generators obtained from \p{strcnst}
after the permutation $2\eta \leftrightarrows \tau$. This
yields the following parametrization of the tilded structure constants
$\mathbf{\tilde J}_{\alpha\beta}=\frac{\mathbf{\tilde J}_{\beta}-\mathbf{\tilde J}_{\alpha}}
{\mathbf{\tilde J}_{\gamma}}$\,,
$\gamma\neq \alpha,\beta$,  where (cf. \p{strcnst})
\begin{equation}
\mathbf{\tilde J}_{1}=\frac{\theta_2(\tau|2\eta)\theta_2(0|2\eta)}
{\theta_2^2({\textstyle \frac{\tau}{2}}|2\eta)}\,,\quad
\mathbf{\tilde J}_{2}=\frac{\theta_3(\tau|2\eta)\theta_3(0|2\eta)}
{\theta_3^2({\textstyle \frac{\tau}{2}}|2\eta)}\,,\quad
\mathbf{\tilde J}_{3}= \frac{\theta_4(\tau|2\eta)\theta_4(0|2\eta)}
{\theta_4^2({\textstyle \frac{\tau}{2}}|2\eta)}\,.
\label{uni2}\end{equation}
The cross-commutation relations between the generators of two Sklyanin algebras,
$\mathbf{S}^\alpha$ and $\mathbf{\tilde S}^\alpha$, have
the form
\begin{eqnarray} \nonumber
&& \mathbf{S}^\alpha \,\mathbf{\tilde S}^\beta=\mathbf{\tilde S}^\beta \,\mathbf{S}^\alpha,
\quad \alpha,\beta\in\{0,3\} \quad
\text{or} \quad \alpha,\beta\in\{1,2\},
\\
&& \mathbf{S}^\alpha \,\mathbf{\tilde S}^\beta=-\mathbf{\tilde S}^\beta \,\mathbf{S}^\alpha,
\quad \alpha\in\{0,3\},\; \beta\in\{1,2\} \quad
\text{or} \quad \alpha\in\{1,2\},\;\beta\in\{0,3\}\,. \lb{cross}
\end{eqnarray}
Because of the non-commutativity of $\mathbf{S}^\alpha$
and $\mathbf{\tilde S}^\beta$ it is not a direct product of two Sklyanin
algebras, though it is not difficult to trace the difference
of actions of the subalgebra generators on modules in different orders.

The Casimir operators of the elliptic modular double second half (cf. \p{Cas1})
\be \lb{Cas2}
\mathbf{\tilde K}_0 = \sum_{\alpha=0}^3\,\mathbf{\tilde S}^\alpha\,\mathbf{\tilde S}^\alpha\ ,\qquad
\mathbf{\tilde K}_2 = \sum_{\alpha=1}^3\,\mathbf{\tilde J}_\alpha\,\mathbf{\tilde S}^\alpha\,
\mathbf{\tilde S}^\alpha\,,
\ee
commute with all generators in this sector:
$[\,\mathbf{\tilde K}_0 ,\mathbf{\tilde S}^\alpha\,] =
[\,\mathbf{\tilde K}_2 ,\mathbf{\tilde S}^\alpha\,] =0$\,.
Then from the cross-commutation relations \p{cross}
it follows that $\mathbf{K}_0, \mathbf{K}_2$ commute with
$\mathbf{\tilde S}^a$ and, vice versa,
$\mathbf{\tilde K}_0,\mathbf{\tilde K}_2$ commute with $\mathbf{S}^\alpha$\,,
i.e. there are in total four Casimir operators:
$$
[\mathbf{K}_0,\mathbf{\tilde S}^\alpha]=[\mathbf{K}_2,\mathbf{\tilde S}^\alpha]=
[\mathbf{\tilde K}_0, \mathbf{S}^\alpha]=[\mathbf{\tilde K}_2, \mathbf{S}^\alpha]=0\,.
$$

For building representations of the described elliptic modular double
the Sklyanin realization of the generators \p{Sklyan1} plays a crucial role.
However, it is convenient in what follows to modify slightly the generators
\p{Sklyan1} by means of a similarity transformation:
$$
\mathbf{S}^\alpha_{mod} \equiv e^{\pi\textup{i}z^2/\eta} \,\mathbf{S}^\alpha\,
e^{-\pi\textup{i}z^2/\eta}.
$$
The reason for such a conjugation is explained in \cite{DS} -- it simplifies the
form of the intertwining operator to be described below.
Obviously, this modification does not alter the Sklyanin's algebra commutation relations \p{SklAlg}.
Since in the rest of the paper we deal only with the new set of generators, we will
omit the mark $``mod"$
for the sake of brevity and denote by $\mathbf{S}^\alpha$ the new set of generators,
\begin{equation}\label{SklyanMod}
\mathbf{S}^\alpha= e^{\pi\textup{i}z^2/\eta}\frac{\textup{i}^{\delta_{\alpha,2}}
\theta_{\alpha+1}(\eta|\tau)}{\theta_1(2 z|\tau) } \Bigl[\,\theta_{\alpha+1} \left(2
z-g +\eta\big|\tau\right)e^{\eta\partial_z} - \theta_{\alpha+1}
\left(-2z-g+\eta\big|\tau\right)e^{-\eta\partial_z}\Bigl]e^{-\pi\textup{i}z^2/\eta}.
\end{equation}
The parameter $g \in \mathbb{C}$ is related to the spin $\ell$ from \p{Sklyan1} as
$g=\eta(2\ell+1)$ and we call it as the {\it spin} as well.
In the following we will use $g$-spin as an independent parameter instead of $\ell$.
The operators \p{SklyanMod} provide a realization for the first Sklyanin algebra of the elliptic modular double.
Permutation of the quasiperiods $2\eta \leftrightarrows \tau$ yields
the finite-difference operator
realization of generators forming the second Sklyanin algebra,
\begin{eqnarray}\nonumber &&
\mathbf{\tilde S}^\alpha  = e^{2\pi\textup{i}z^2/\tau}
\frac{\textup{i}^{\delta_{\alpha,2}}\theta_{\alpha+1}({\textstyle \frac{\tau}{2}}|2\eta)}{\theta_1(2 z|2\eta) }
 \Bigl[\,\theta_{\alpha+1} \left(2z-g+{\textstyle \frac{\tau}{2}}\big|2\eta\right)
 \mathrm{e}^{\frac{1}{2}\tau \partial_z} -
\\ && \makebox[10em]{}
 - \theta_{\alpha+1}\left(-2z-g+ {\textstyle \frac{\tau}{2}}\big|2\eta\right)
\mathrm{e}^{-\frac{1}{2}\tau \partial_z}\, \Bigl]e^{-2\pi\textup{i}z^2/\tau},
\label{mod_doub2}\end{eqnarray}
where the $g$-spin is the same arbitrary parameter as in \eqref{SklyanMod}.
The cross-commutation relations \p{cross} can be verified for the set
of generators \p{SklyanMod} and \p{mod_doub2}.

In the considered realization of the elliptic modular double four Casimir operators
\p{Cas1} and  \p{Cas2} reduce to the following
scalar expressions
\begin{eqnarray*} &&
\mathbf{K}_0 = 4\,\theta_1^2\bigl(g|\tau\bigr)\ ,\quad
\mathbf{K}_2 = 4\,\theta_1\bigl(g-\eta|\tau\bigr)\,
\theta_1(g+\eta|\tau)\,,
\\ &&
\mathbf{\tilde K}_0 = 4\,\theta_1^2\bigl(g|2\eta\bigr)\ ,\quad
\mathbf{\tilde K}_2 = 4\,\theta_1\bigl(g-{\textstyle \frac{\tau}{2}}|2\eta\bigr)\,
\theta_1(g+{\textstyle \frac{\tau}{2}}|2\eta)\,,
\end{eqnarray*}
which  are invariant under the reflection $g\to -g$.
The variables $\eta$ and $\tau$ are fixed by the structure constants \p{strcnst}
(or \p{uni2}). Therefore the $g$-spin parameter fixes the values of
all Casimirs and specifies representations of the elliptic modular double.

In the previous section we considered the Lax operator for the Sklyanin algebra.
Let us extend corresponding formulae to the elliptic modular double.
Since the latter contains two halves there are two different L-operators $\mL^{doub}$
and $\widetilde{\mL}^{doub}$. The L-operator related to the first algebra is
constructed out of the generators $\mathbf{S}^\alpha$, i.e. $\mL^{doub}$ is
given by the expression \p{L_op} with $\mathbf{S}^\alpha$ fixed in \p{SklyanMod},
\be \lb{Lax1}
\mathrm{L}^{doub}(u)=
\sum_{\alpha=0}^3 w_{\alpha} (u)\, \sigma_\alpha \otimes
\mathbf{S}^\alpha, \qquad
w_{\alpha}(u) = \frac{\theta_{\alpha+1}(u+\eta|\tau)}{\theta_{\alpha+1}(\eta|\tau)}\,.
\ee
Analogously to \p{LFact} it can be factorized as follows
\begin{align}\notag
\mathrm{L}^{doub}(u_1,u_2) = &e^{\pi\textup{i}z^2/\eta}\,
\frac{1}{\theta_1(2 z|\tau)} \left(
\begin{array}{cc}
\theta_3\left(z - u_1|\frac{\tau}{2}\right) &
-\theta_3\left(z+u_1|\frac{\tau}{2}\right) \\
-\theta_4\left(z - u_1|\frac{\tau}{2}\right) & \theta_4\left(z+u_1|\frac{\tau}{2}\right)
\end{array} \right)
\left(
\begin{array}{cc}
\mathrm{e}^{\eta\partial} &0\\
0 & \mathrm{e}^{-\eta\partial}
\end{array} \right)\cdot
\\
&\cdot\left(
\begin{array}{cc}
\theta_4\left(z+u_2|\frac{\tau}{2}\right) &
\theta_3\left(z+u_2|\frac{\tau}{2}\right) \\
\theta_4\left(z - u_2|\frac{\tau}{2}\right) &
\theta_3\left(z - u_2|\frac{\tau}{2}\right)
\end{array} \right)\,e^{-\pi\textup{i}z^2/\eta}\,, \label{doub}
\end{align}
where the ``light-cone" combinations of the spectral parameter
and $g$-spin are (cf. \p{u1u2})
\be \lb{u12doub}
u_1 =
\frac{u+g}{2}\ \,,\
u_2 = \frac{u-g}{2}\,.
\ee
Similarly, the L-operator for the second Sklyanin algebra is constructed
out of the generators $\mathbf{\tilde S}^\alpha$ \p{mod_doub2},
\be \lb{Lax2}
\widetilde{\mathrm{L}}^{doub}(u)=
\sum_{\alpha=0}^3 \widetilde{w}_{\alpha} (u)\, \sigma_\alpha \otimes
\widetilde{\mathbf{S}}^\alpha, \qquad
\widetilde{w}_{\alpha}(u) = \frac{\theta_{\alpha+1}(u+{\textstyle \frac{\tau}{2}}|2\eta)}{\theta_{\alpha+1}({\textstyle \frac{\tau}{2}}|2\eta)}\,.
\ee
It takes the factorized form as well which is obtained from \p{doub} by
interchange of the quasiperiods $\tau\rightleftarrows 2\eta$.
Both Lax operators \p{Lax1} and \p{Lax2} respect YBE \p{Lax} with Baxter's R-matrices.
For $\mL^{doub}$ it is the R-matrix \p{RBaxter}, and for $\widetilde{\mL}^{doub}$ the R-matrix is
given by \p{RBaxter} with $2\eta \leftrightarrows \tau$
(i.e. by \p{RBaxter} with the weights $w_{\alpha} (u)$ substituted for $\widetilde{w}_{\alpha} (u)$).

\subsection{The intertwining operator}

Since we are going to construct finite-dimensional solutions of YBE \p{YB} with the symmetry of the elliptic
modular double,
we need to describe the structure of its finite-dimensional representations.
It is well known that intertwining operators of equivalent representations provide
an extremely useful tool enabling one to uncover finite-dimensional representations.
They give an evidence on the decoupling of finite-dimensional representations
from infinite-dimensional ones through their null-spaces.

Let us consider the following integral operator $\mathrm{M}(g)$ depending on a parameter
$g \in \mathbb{C}$ acting on some function of one complex variable $\Phi(z)$ as
\begin{equation}
[\,\mathrm{M}(g)\,\Phi\,](z)= \kappa
\int_0^1 \frac{\Gamma(\pm z\pm x -g)}
{\Gamma(-2g,\pm 2x)}\Phi(x)\,dx,
\label{M}\end{equation}
where  we denoted
$$
\kappa = \frac{1}{2}(q;q)_\infty\,(p;p)_\infty, \quad
p=e^{2\pi\textup{i}\tau}, \quad q=e^{4\pi\textup{i}\eta}.
$$
Parameters $p$ and $q$ are quasiperiods in the multiplicative notation
(cf. \p{multnot}).
In order to avoid bulky expressions we extensively use the shorthand converntions
\begin{eqnarray*} &&
\Gamma(a,b) :=\Gamma(a)\Gamma(b), \qquad \Gamma(\pm x):= \Gamma(x)\Gamma(-x),
\\ &&
\Gamma(\pm z \pm x):= \Gamma(z+x) \Gamma(z-x) \Gamma(-z+x) \Gamma(-z-x)
\end{eqnarray*}
for products of the elliptic gamma function \cite{FV,rui,AA2003}
\begin{equation}
\Gamma(z)\equiv\Gamma(z|\tau,2\eta):=
\prod_{n,m=0}^{\infty} \frac{1-\mathrm{e}^{-2\pi\textup{i}z}
p^{n+1}q^{m+1}}{1-\mathrm{e}^{2\pi\textup{i}z}
p^ nq^m}
\label{egamma}\ee
defined for $|p|,|q|<1$.
This special function is omnipresent in our considerations and
it possesses a number of important identities indicated below.
The $\Gamma(z)$-function satisfies functional equations that justify its name,
\begin{align}
&\Gamma(z+2\eta) = \mathrm{R}(\tau)\,e^{\textup{i}\pi z}\,\theta_1(z|\tau)\,\Gamma(z)\,, \notag\\
&\Gamma(z+\tau) = \mathrm{R}(2\eta)\,e^{\textup{i}\pi z}\,
\theta_1(z|2\eta)\,\Gamma(z)\,,
\lb{Gshift}
\end{align}
where on the right-hand sides one actually has the coefficients
$\theta(e^{2\pi i z};p)$ and $\theta(e^{2\pi i z};q)$, respectively
(cf. \eqref{theta1} and \p{multnot}).
One has the reflection identity
\be \lb{refl}
\Gamma(z) \,\Gamma(-z + 2\eta +\tau) = 1\,.
\ee
Zeros of $\Gamma(z)$ are located on the two-dimensional lattice
$z = \mathbb{Z} + \tau\mathbb{Z}_{>0} + 2 \eta \mathbb{Z}_{>0}$
and poles on the lattice $z = \mathbb{Z} + \tau \mathbb{Z}_{\leq 0} + 2\eta\mathbb{Z}_{\leq 0}$.

The integral operator $\mathrm{M}(g)$ was introduced in \cite{spi:bailey}
in order to define a universal integral transform of hypergeometric type
yielding an integral analogue of the Bailey chain techniques \cite{aar}.
It is relevant to YBE since the general $\mR$-operator factors as a product of two
operators of the form \eqref{M} intertwined by two explicit functions
similar to the integrand of $\mathrm{M}(g)$ (see \cite{DS} and Sect. \ref{SecR}).

Let us discuss the domain of parameter values and the space of functions for which
the formal expression for $\mathrm{M}(g)$ becomes a rigorously defined
operator with the properties needed for our considerations below.
First we introduce the multiplicative variables
$$
Z:=e^{2\pi \textup{i} z}, \quad X:=e^{2\pi \textup{i} x}, \quad t:=e^{-2\pi \textup{i} g}, \quad
\Gamma(Z;p,q):=\Gamma(z|\tau,2\eta),
$$
and demand that we work in the space of periodic analytical functions
$\Phi(x+1)=\Phi(x)$, which allows us to pass from $\Phi(x)$ to analytical
functions of $X\in\C^*$, $f(X):=\Phi(x)$. So, we can now write
\begin{equation}
[\mathrm{M}(g)f](Z)= \kappa
\int_{\mathbb{T}} \frac{\Gamma(tZ^{\pm1}X^{\pm1};p,q)}
{\Gamma(t^2,X^{\pm 2};p,q)}f(X)\frac{dX}{2\pi \textup{i}X},
\label{M_mult}\end{equation}
where $\mathbb{T}$ is the unit circle of positive orientation and
$$
\Gamma(a,bc^{\pm k};p,q):=\Gamma(a;p,q)\Gamma(bc^k;p,q)\Gamma(bc^{-k};p,q).
$$
The requirement of analyticity in $Z$ is natural since after the action of
the operator $\mathrm{M}(g)$ we obtain such functions (other explicit
functions involved in our considerations will also satisfy this constraint).

Next we require that $|tZ^{\pm 1}|<1$ (or Im$(-g\pm z)>0$) and suppose that
$f(X)$ are holomorphic functions of $X$. Under these conditions no singularity
lies on the integration contour $\T$ which separates geometric progressions
of poles of the integrand in \eqref{M_mult} converging to $Z=0$ from sequences
of poles going to infinity. These constraints completely remove ambiguities in
the definition of $\mathrm{M}(g)$. However, for our purposes we have to define
this integral operator on a wider domain of values of parameters
and impose additional constraints on the test functions  $\Phi(x)$ (or $f(X)$).
We can easily continue analytically the definition of  $\mathrm{M}(g)$ to all values of
$t\in\C^*$ (or, $g\in \C$). Indeed, let us pull $t$ and $Z$
away from the domain indicated above and simultaneously deform $\T$ to a contour $C$
such that it still separates converging and diverging sequences of poles of
the integrand. This procedure yields the necessary analytical continuation
of $\mathrm{M}(g)$. Evidently, we can contract $C$ back to $\T$ and, in case if we
did not put occasionally a singularity on $\T$, represent $\mathrm{M}(g)$ as a sum
of the integral operator \eqref{M_mult} (with the integration contour $\T$) and
the operator of the form $[\mathrm{M}^{res}(g)f](Z):=\sum_j c_j(Z)f(X_j)$, where
$c_j(Z)$ are residues of the integrand poles with the positions $X=X_j$ that
are crossed over during the contraction of $C$ to $\T$.

However, for some particular values of parameters the required contour $C$ may not exist,
which happens when a number of poles pinch the integration contour from two sides.
Such a situation was considered in detail in papers \cite{DS,DS2} and we describe briefly
their conclusions. Denote $t_1:=tZ$ and $t_2:=tZ^{-1}$. Then there are two possibilities.
In the first case $t_1t_2=t^2 \to q^{-n}p^{-m}$ (i.e., $g\to n\eta+m\tau/2$ or $g\to 1/2+ n\eta+m\tau/2$),
$n,m\in\Z_{\geq 0}$ and $t_1\neq t_2$. Then $(n+1)(m+1)$ poles start to pinch the
integration contour. After crossing over the half of these poles by the integration
contour, picking up the residues and taking the limiting values of $t$, the integral
operator $\mathrm{M}(n\eta+m\tau/2)$ is converted to a finite difference operator
with the coefficients composed out of the Jacobi theta-functions of $z$.
In the second case, one has either $t_1^2=q^{-n_1}p^{-m_1}$ and/or $t_2^2=q^{-n_2}p^{-m_2}$,
$n_{1,2},m_{1,2}\in\Z_{\geq 0}$. However, in the latter case we have constraints on
the variable $Z$, which breaks analytical structure of our functions. Demanding that our
operator  $\mathrm{M}(g)$ maps analytical functions $f_{in}(X)$ to analytical functions
$f_{out}(Z)$, we are left with the first option with the constraint
$Z^{\pm 2}\neq t^{2}q^{j}p^{k},\, j,k\in\Z_{\geq 0}$ (where one has singularities
of the coefficients of the finite-difference operator $\mathrm{M}(n\eta+m\tau/2)$).

Still, these are not all constraints we need. Until now we were assuming that
our test functions are holomorphic. If they are meromorphic, then it is necessary
to repeat the analytical continuation procedure by taking into account the
singularities sitting in the test functions.
In the following we will act by different $\mathrm{M}(g)$-operators on different
meromorphic functions. However, we shall not describe explicitly the appropriate
integration contours satisfying all necessary conditions taking the assumption that
they do exist. The final key result -- construction of new finite-dimensional
R-matrices and L-operators will be confirmed by an independent derivation of
the same result using a completely different approach.

It was rigorously shown in \cite{spi-war:inversions}
that for generic values of $g$ and $z$ from some restricted domain our integral operator
satisfies a very simple inversion relation resembling the key Fourier transformation property
\begin{equation}
\mathrm{M}(g)\, \mathrm{M}(-g) = \II\,,
\label{inv}\end{equation}
provided the operators act in the space of even functions, $\Phi(-x)=\Phi(x)$
(or, equivalently, $f(X^{-1})=f(X)$), which are holomorphic in a bounded domain.
In \eqref{inv} it is assumed that at least
one of the operators $\mathrm{M}(\pm g)$ is defined by the analytical
continuation since it is not possible to satisfy simultaneously the constraints
Im$(\pm g\pm z)>0$ (or, $|t^{\pm}Z^{\pm 1}|<1$) for all  choices of the signs.
In order to preserve this important inversion property, we demand that all our test functions
are even. Then, for the spin values away from two discrete
lattices $\pm g =n \eta + m \frac{\tau}{2}$ and $\pm g = \frac{1}{2} + n \eta + m \frac{\tau}{2}$,
we can confirm the inversion relation \eqref{inv} in a domain allowed
by the integration contours with appropriate properties for involved integral operators.
Namely, the integration contour of $\mathrm{M}(-g)$  should contain the
points $t^{-1}Z^{\pm1}p^jq^k$ inside it and exclude their reciprocals, whereas
the integration contour in $Z$ for $\mathrm{M}(g)$ should separate points $t^{-1}Y^{\pm1}p^jq^k$
and their reciprocals ($Y$ is an external variable for $\mathrm{M}(g)$) under the restriction
$Z^{\pm2}\neq t^{-2}q^{j}p^{k},\, j,k\in\Z_{\geq 0}$.

In \cite{DS} it was proved that for some mild restrictions on the parameters the operator \eqref{M},
being symmetric in $2\eta$ and $\tau$, satisfies the following intertwining relations with
the generators of the elliptic modular double,
\begin{equation}
\mathrm{M}(g)\,\mathbf{S}^\alpha(g) =
\mathbf{S}^\alpha(-g)\, \mathrm{M}(g), \qquad
\mathrm{M}(g)\,\mathbf{\tilde S}^\alpha(g) =
\mathbf{\tilde S}^\alpha(-g)\,\mathrm{M}(g)\,.
\label{inter1}
\end{equation}
Here we explicitly indicate the $g$-spin dependence of the algebra generators
(see \p{SklyanMod} and \p{mod_doub2}) in order to show that $g$ changes the
sign under the action of $\mathrm{M}$. Thus the integral operator
$\mathrm{M}(g)$ \p{M} is the intertwining operator of equivalent representations
of the elliptic modular double \cite{DS}. Evidently, it is
an intertwining operator of equivalent representations of the Sklyanin algebra itself.
However, in the single Sklyanin algebra case there is a family of intertwining operators
depending on a periodic function of two complex variables, i.e. there is a functional
freedom in their choice.
The elliptic modular double enables one to fix uniquely the representative \p{M} out of
this family.
In the conventional Sklyanin algebra setting usage of the spin variable $\ell$
is preferable. From the relation $g=\eta(2\ell+1)$ one can see that the
equivalent representations are obtained by the transformation $\ell\to -1-\ell$.

Now we consider the intertwining operator \p{M} for the two-index discrete lattice
of the $g$-spin parameter
$g = n\eta+m\tau/2$ with $n,m \in \mathbb{Z}_{\geq 0}$\,.
In this case $\mathrm{M}(g)$ drastically simplifies.
This can be seen by means of the
contiguous (or recurrence) relations for the intertwining operator \cite{CDKKIII,DS2}
\be \lb{cont}
\mathrm{A}_k(g)\,\mathrm{M}(g) := \mathrm{M}(g+\eta)\,\theta_k
 \left(z | {\textstyle\frac{\tau}{2}}\right),\qquad
\mathrm{B}_k(g)\,\mathrm{M}(g) := \mathrm{M}\left(g+{\textstyle\frac{\tau}{2}}\right)\,
\theta_k \left(z | \eta\right)\,,
\ee
where $\mathrm{A}_k(g)$ and $\mathrm{B}_k(g)$, $k=3,4$, are the following
difference operators
\begin{align*}
&\mathrm{A}_k(g) = e^{\pi \textup{i}\frac{z^2}{ \eta}}\,\frac{c_A}{\theta_1(2z | \tau)}
\left[ \theta_k \left(z+g+\eta | {\textstyle\frac{\tau}{2}}\right)\, e^{\eta \partial_z} -
\theta_k \left(z-g-\eta | {\textstyle\frac{\tau}{2}}\right)\, e^{-\eta \partial_z}
\right]\, e^{-\pi \textup{i}\frac{z^2}{ \eta}}\,,
\\
&\mathrm{B}_k(g) = e^{2\pi \textup{i}\frac{z^2}{ \tau}}
\,\frac{c_B}{\theta_1(2z | 2\eta)}
\left[ \theta_k \left(z+g+{\textstyle\frac{\tau}{2}}| \eta\right)
\, e^{{\textstyle\frac{\tau}{2}} \partial_z} -
\theta_k \left(z-g-{\textstyle\frac{\tau}{2}}| \eta\right)
\, e^{-{\textstyle\frac{\tau}{2}} \partial_z}
\right]\, e^{-2\pi \textup{i}\frac{z^2}{ \tau}}\,,
\end{align*}
with the normalization constants
$$
c_A = \frac{e^{\pi \textup{i} \eta}}{\mathrm{R}(\tau)}\ \ \,,\ \
c_B = \frac{e^{\pi \textup{i} {\textstyle\frac{\tau}{2}}}}{\mathrm{R}(2\eta)}\,.
$$

The contiguous relations \p{cont} describe transformations of $\mathrm{M}(g)$ under shifts along the two-dimensional
$g$-spin lattice $g = n\eta+m\tau/2$, $n,m\in\Z_{\geq0}$.
As shown in \cite{DS2} (see also \cite{CDKKIII}) they
lead to a factorized representation of $\mathrm{M}(g)$ on this lattice.
Indeed, using the initial condition $\mathrm{M}(0) = \II$, which is proved by the
residue calculus \cite{DS}, one solves straightforwardly the contiguous relations \p{cont}
in $\eta$-direction or $\frac{\tau}{2}$-direction on the lattice
and obtains
\begin{align} \lb{Meta}
&\mathrm{M}(n\eta) = \mathrm{A}_k(n\eta-\eta)\cdots \mathrm{A}_k(\eta) \mathrm{A}_k(0)\cdot
\theta_k^{-n} \left(z | {\textstyle\frac{\tau}{2}}\right)\,,
\\
&\mathrm{M}\left( m {\textstyle\frac{\tau}{2}}\right) = \mathrm{B}_k\left(m {\textstyle\frac{\tau}{2}} -{\textstyle\frac{\tau}{2}}\right)\cdots \mathrm{B}_k\left({\textstyle\frac{\tau}{2}}\right) \mathrm{B}_k(0)\cdot
\theta_k^{-m} \left(z | \eta\right)\,. \notag
\end{align}
Note that the form of the intertwiner does not depend on the values of index $k$ for involved
theta functions.
The general operator $\mathrm{M}\left(n\eta + m {\textstyle\frac{\tau}{2}}\right)$
has the form~\cite{DS2}
\begin{eqnarray}\nonumber  &&
\mathrm{M}\left(n\eta + m {\textstyle\frac{\tau}{2}}\right) =
\mathrm{A}_k(n\eta-\eta+ m {\textstyle\frac{\tau}{2}})\cdots
\mathrm{A}_k(\eta+m {\textstyle\frac{\tau}{2}})
\mathrm{A}_k(m {\textstyle\frac{\tau}{2}})\,\cdot
\\  && \makebox[2em]{}
\cdot\,
\mathrm{B}_k\left(m {\textstyle\frac{\tau}{2}} -{\textstyle\frac{\tau}{2}}\right)\cdots \mathrm{B}_k\left({\textstyle\frac{\tau}{2}}\right) \mathrm{B}_k(0)
\cdot
\theta_k^{-m} \left(z | \eta\right)
\theta_k^{-n} \left(z | {\textstyle\frac{\tau}{2}}\right)\,.
\label{genform}\end{eqnarray}
Thus we see that $\mathrm{M}\left(n\eta + m {\textstyle\frac{\tau}{2}}\right)$
is a finite-difference operator with elliptic coefficients.
Of course there are many equivalent ways to represent
$\mathrm{M}\left(n\eta + m {\textstyle\frac{\tau}{2}}\right)$ as a product of
$\mathrm{A}_k$- and $\mathrm{B}_k$-operators corresponding to all possible zigzags on the lattice
from the point $g = 0$ to $g = n \eta + m \tau/2$. In \p{genform} we choose the trajectory
starting from $g=0$, going along the $\frac{\tau}{2}$-direction to $g= m\tau/2$, then turning
towards the $\eta$-direction and proceeding to $g = n \eta + m \tau/2$.

Expanding \p{genform} one obtains the intertwiner in the form of a sum
over shift operators $e^{(k\eta+ l \frac{\tau}{2})\dd}$\,, $k, l\in \mathbb{Z}$,
with elliptic coefficients. Explicit expressions can be found in \cite{DS2}.
In this form the discrete intertwining
operator $\mathrm{M}(n \eta)$ has been obtained first in \cite{Z}.

There is the second lattice $g = \frac{1}{2} + n \eta + m \frac{\tau}{2}$, $n,m \in \mathbb{Z}_{\geq 0}$\,,
leading to a crucial simplification of the intertwiner.
The residue calculation from \cite{DS} shows that the intertwiner \p{M} simplifies
at $g=\frac{1}{2}$ to
$\mathrm{M}(\frac{1}{2}) = e^{\frac{1}{2}\dd}$.
As shown in \cite{DS2}, the intertwiners on these two lattices are related to each other
 by the half-period shift operator
\be \lb{M1/2}
\textstyle\mathrm{M}(\frac{1}{2} + n \eta + m  \frac{\tau}{2}) =
\mathrm{M}(n \eta + m \frac{\tau}{2}) \,e^{\frac{1}{2}\dd}\,.
\ee

\subsection{Finite-dimensional representations of the elliptic modular double}
\lb{SecFD}

The intertwining operator gives us an insight to finite-dimensional representations
of the elliptic modular double.
Indeed, equalities \eqref{inter1} show that the null-space of the operator
$\mathrm{M}(g)$ and the image of the operator $\mathrm{M}(-g)$ form
invariant spaces for the elliptic modular double, i.e. they both are
invariant under the action of two constituent Sklyanin algebra
generators ${\mathbf S}^\alpha(g)$ \eqref{SklyanMod}
and  $\mathbf{\tilde S}^\alpha(g)$ \eqref{mod_doub2},
\be \lb{invspace}
{\mathbf S}^\alpha,\,\mathbf{\tilde S}^\alpha\,: \mathrm{Ker}\, \mathrm{M}(g) \to \mathrm{Ker}\, \mathrm{M}(g)
,\qquad
{\mathbf S}^\alpha,\,\mathbf{\tilde S}^\alpha\,: \mathrm{Im}\,\mathrm{M}(-g) \to \mathrm{Im}\,\mathrm{M}(-g)\,.
\ee
Consequently, if the intertwining operator has a nontrivial null-space then
a sub-representation of the algebra decouples and one naturally comes to the
corresponding invariant subspace.

The null-space of the intertwiner \p{M} was partially characterized in \cite{DS}
and extensively investigated in \cite{DS2}. There it was proven in two ways --
by using the factorized representation \p{genform}
and taking the limits of $g$-spin values in the integral representation \p{M} --
that a nontrivial null-space of the intertwiner arises for two
distinct $g$-spin lattices:
$$
g= n \eta + m \frac{\tau}{2} \;\;\;\;\text{and}\;\;\;\;
g = \frac{1}{2} + n \eta + m \frac{\tau}{2} \;, \;\;n,m \in \mathbb{Z}_{\geq 0}\,,\,(n,m)\neq (0,0)\,.
$$
Here we outline the second proof since it provides a natural basis
of finite-dimensional representations and the corresponding generating function.

The whole null-space of the intertwiner is too big, so
we need imposing some additional constraints.
The intersection of two invariant subspaces \p{invspace} is a finite-dimensional
representation of the elliptic modular double~\cite{DS2}
\be \lb{KerIm}
\mathrm{Ker}\, \mathrm{M}(g)\cap \mathrm{Im}\,\mathrm{M}_{ren}(-g),
\ee
where the $g$-spins belong to the lattices
\be \lb{lattice}
g=(n+1) \eta + (m+1) \frac{\tau}{2} \;\;\;\;\text{or}\;\;\;\;
g = \frac{1}{2} + (n+1) \eta + (m+1) \frac{\tau}{2} \;, \;\;n,m \in \mathbb{Z}_{\geq 0}\,,
\ee
and $\mathrm{M}_{ren}(g)$ is the renormalized intertwining operator
which, obviously, fulfills the intertwining relations \p{inter1} with the
elliptic modular double generators (cf. \p{M}),
\begin{equation}
[\,\mathrm{M}_{ren}(g)\,\Phi\,](z)= \kappa
\int_0^1 \frac{\Gamma(\pm z\pm x -g)}
{\Gamma(\pm 2x)}\Phi(x)\,dx\,.
\label{M_ren}
\end{equation}

Using the recurrence relations \eqref{cont}, in \cite{DS2} it was shown by
direct computation that for $g$ lying on the lattices \eqref{lattice} the intertwiner
$\mathrm{M}(g)$ annihilates
the integral kernel of the operator $\mathrm{M}_{ren}(-g)$ \p{M_ren}
considered as a function of $z$ and containing an auxiliary parameter $x$, e.g.
$$
\mathrm{M}_z\left((n+1)\eta+(m+1)\textstyle{\frac{\tau}{2}}\right)\cdot
\,\Gamma\left(\pm  z \pm  x + (n +1)\eta + (m+1) {\textstyle\frac{\tau}{2}}\right) = 0\,.
$$
Here the subindex $z$ indicates that the $\mathrm{M}_z$-operator acts by integration
over the $z$-variable and we denote the ``external" variable by the same letter $z$.
This result would follow directly from the inversion relation \eqref{inv}, provided
one continues analytically both operators to $g$ values lying near
the needed discrete lattice points, which requires search of appropriate
contours of integration. Then one simply multiplies \eqref{inv}
by the numerical factor $\Gamma(2g)$ and take the limit
$g\to (n+1)\eta+(m+1)\textstyle{\frac{\tau}{2}}$
at $n,m\in\Z_{\geq 0}$.
Since $\Gamma(2g)$ vanishes in this limit (recall \p{egamma}) we come to
the operator identity
\begin{equation}
\mathrm{M}\left((n+1)\eta+(m+1)\textstyle{\frac{\tau}{2}}\right)\,
\mathrm{M}_{ren}\left(-(n+1)\eta-(m+1){\textstyle\frac{\tau}{2}}\right)=0\,.
\label{zero1}
\end{equation}
This relation should hold after application to an arbitrary test function, therefore
the integrand function of $\mathrm{M}_{ren}(-g)$ must belong to the null space of $\mathrm{M}(g)$
for $g$ lying on the lattices \p{lattice}. We conclude that the function
\be \lb{genfun1}
K(z,x):=\Gamma\left(\pm  z \pm  x + (n +1)\eta + (m+1) {\textstyle\frac{\tau}{2}}\right),
\quad n,m\geq 0,
\ee
is the generating function of $(n+1)(m+1)$-dimensional
representation for the first spin lattice in \eqref{lattice}.
Expanding it with respect to theta-functions
of the auxiliary parameter $x$ we recover all basis vectors of the finite-dimensional
representation.
Because the generating function \p{genfun1} coincides with the integrand of $\mathrm{M}_{ren}$,
all basis vectors belong not only to the null-space of $\mathrm{M}(g)$,
but they also lie in the intersection with Im$\,\mathrm{M}_{ren}(-g)$ \p{KerIm}.
Moreover, this representation is irreducible.

As to the boundary values of the spin $g=n\eta,\, n>0,$ or $g=m\tau/2,\, m>0,$
we obtain the null subspace, but it is infinte-dimensional. As shown in \cite{DS2},
in this case only one of the Sklyanin subalgebras of the elliptic modular double
has an $n$- or $m$-dimensional representation, but its partner algebra representation
appears to be infinite-dimensional.

Now we explicitly extract basis vectors from \p{genfun1}.
We simplify the generating function to a finite product of theta functions
using equations \p{Gshift} and the inversion relation \p{refl},
\begin{eqnarray*} &&
K(z,x)
=\frac{\Gamma\left(x \pm  z + (1+n)\eta + (1+m) {\textstyle\frac{\tau}{2}}\right)}
{\Gamma\left(x \pm  z + (1-n)\eta + (1-m) {\textstyle\frac{\tau}{2}}\right)} =
\\ && \makebox[2em]{}
=\prod_{r=0}^{n-1}\theta(e^{2\pi i(x \pm  z + (2r+1-n)\eta + (1+m)
{\textstyle\frac{\tau}{2}})};p)\prod_{s=0}^{m-1}\theta(e^{2\pi i(x \pm  z + (1-n)\eta
+ (2s+1-m){\textstyle\frac{\tau}{2}})};q).
\end{eqnarray*}
Passing to $\theta_{3,4}(z|{\textstyle\frac{\tau}{2}})$
and $\theta_{3,4}(z|\eta)$ functions, one can rewrite the latter products as
\begin{align}
c_{nm} \,\cdot &
\sideset{}{_{r= 0}^{n-1}}\prod
\left[ \,\theta_3(z|{\textstyle\frac{\tau}{2}}) \,\theta_4\left(x+(n-1-2r)\eta\,\big|{\textstyle\frac{\tau}{2}}\right)
+ (-1)^m \,\theta_4(z|{\textstyle\frac{\tau}{2}}) \,\theta_3\left(x+(n-1-2r)\eta\,\big|{\textstyle\frac{\tau}{2}}\right) \,\right]
\cdot \notag
\\ \cdot &
\sideset{}{_{s = 0}^{m-1}}\prod
\left[ \,\theta_3(z|\eta) \,\theta_4\left(x+(m-1-2s){\textstyle\frac{\tau}{2}}\,\big|\eta\right)
+ (-1)^n \,\theta_4(z|\eta) \,\theta_3\left(x+(m-1-2s){\textstyle\frac{\tau}{2}}\,\big|\eta\right)
\,\right]\, ,
\lb{genfunellip}
\end{align}
where $c_{nm}$ is a normalization constant,
$$
c_{nm}= (-2)^{-m-n} \mathrm{R}^{2n}(\tau) \,\mathrm{R}^{2m}(2\eta)
e^{\pi i[n(1-m^2)\frac{\tau}{2}+ m(1-n^2)\eta ]}.
$$
Thus we conclude that the set of $(n+1)(m+1)$ homogeneous
polynomials in theta functions of degree $n$ with respect to
$\theta_3 \left(z | {\textstyle\frac{\tau}{2}}\right)$ and
$\theta_4 \left(z | {\textstyle\frac{\tau}{2}}\right)$ and
of degree $m$ with respect to
$\theta_3 \left(z | \eta \right)$ and
$\theta_4 \left(z | \eta \right)$, i.e. the functions
\begin{equation}\label{phi}
\varphi_{j,l}^{(n,m)}(z) =
\left[\theta_3 \left(z | {\textstyle\frac{\tau}{2}}\right)\right]^j \,
\left[\theta_4 \left(z | {\textstyle\frac{\tau}{2}}\right)\right]^{n-j}
\cdot
\left[\,\theta_3 \left(z | \eta \right)\,\right]^l \,
\left[\,\theta_4 \left(z | \eta \right)\,\right]^{m-l}
\end{equation}
with $j = 0, 1, 2, \ldots, n$\, and $l = 0, 1, 2, \ldots, m$\,
form a basis of the finite-dimensional representation which
arises at the spin values $g = (n+1) \eta + (m+1) \frac{\tau}{2}$\,, with $n,\,m \in \mathbb{Z}_{\geq 0}$.
This result is in line with the basis \p{basisSkl} in the space
of finite-dimensional representations of the Sklyanin algebra.

Note that the variable $z$ and the auxiliary parameter $x$ appear on equal footing in
the arguments of $\Gamma$-functions in \p{genfun1},
whereas the symmetry $z \rightleftarrows x$ is not obvious in the expression \p{genfunellip}.
Interchanging  $z \rightleftarrows x$ we obtain thus another expansion of the
generating function in theta functions.
Consequently, the generating function \p{genfunellip} produces two natural bases: the basis \p{phi}
$\{ \varphi_{j,l}^{(n,m)}(z) \}^{l = 0,1, \cdots,m}_{j = 0,1, \cdots,n}$
and the basis $\{ \psi_{j,l}^{(n,m)}(z) \}^{l = 0,1, \cdots,m}_{j = 0,1, \cdots,n}$
formed by sums of theta function products with the arguments shifted by a
multiple of $\eta$ or $\frac{\tau}{2}$,
\be \lb{psi}
\psi_{j,l}^{(n,m)}(z) :=  \mathrm{Sym}
\prod_{r=0}^{n-1} \theta_{a_r}\left(z+(n-1-2r)\eta\,\big|{\textstyle\frac{\tau}{2}}\right)
\cdot
\mathrm{Sym}\prod_{s=0}^{m-1} \theta_{b_s}\left(z+(m-1-2s){\textstyle\frac{\tau}{2}}\,\big|\eta\right),
\ee
where $a_r , b_s \in \{3,4\}$, $\theta_3$ appears $j$ times in the first product and
$l$ times in the second product,
and $\mathrm{Sym}$ stands for symmetrization with respect to indices $3,4$ specifying theta functions.
Using these bases we can write
\begin{equation}
K(z,x)= \sum_{j=0}^n\sum_{l=0}^m C_{jl}\, \varphi_{j,l}^{(n,m)}(z)\psi_{n-j,m-l}^{(n,m)}(x)
\label{repr-ker}\end{equation}
for some constants $C_{jl}$. If one introduces a scalar product on the
representation space such that $K(z,x)$ will be the reproducing kernel on it, then
$\varphi_{j,l}^{(n,m)}$ and $\psi_{j,l}^{(n,m)}$ become dual bases.

Analogous results take place for the second $g$-spin lattice.
The generating function of finite-dimensional representations at spin
$g = \frac{1}{2} + (n+1) \eta + (m+1) \frac{\tau}{2}$\,, with $n,m \in \mathbb{Z}_{\geq 0}$, is
\be \lb{genfun2}
\Gamma\left(\pm  z \pm  x + \textstyle\frac{1}{2}+ (n +1)\eta + (m+1) \frac{\tau}{2}\right)\,.
\ee
Since the shift $z \to z+\frac{1}{2}$ results in the permutation of theta functions $\theta_3(z)$ and $\theta_4(z)$,
$$
e^{\frac{1}{2}\dd} \,:\, \left(\theta_3(z),\theta_4(z)\right) \to \left(\theta_4(z),\theta_3(z)\right),
$$
we conclude that $\varphi_{j,l}^{(n,m)}(z)$ \p{phi} and $\psi_{j,l}^{(n,m)}(z)$ \p{psi} still form
the appropriate bases.

\section{The general elliptic $\mathrm{R}$-operator}
\lb{SecR}

\subsection{Intertwining operators and factorization of the $\mathrm{R}$-operator}

We increase once more the complexity level in YBE \p{YB}
and describe its solutions with the symmetry of the elliptic modular
double (see Sect. \ref{SecDoub}), which imposes more severe restrictions as
compared to the plain Sklyanin algebra case.
The reason for this choice will become evident shortly.
The most general known R-operator, i.e. an integral operator solution of the equation
\p{RLL'} with an elliptic hypergeometric kernel,
has been constructed in \cite{DS} following the general ideological scheme
of \cite{Derkachov:2007gr} powered by the techniques of elliptic hypergeometric
integrals \cite{spi:umn,AA2003,spi:essays}.
The construction naturally implies factorization of the R-operator to a product
of several simple operators.
In \cite{Derkachov:2007gr} a number of factorized forms of the R-operator
has been derived which were related to an integral operator realization
of the symmetric group $\mathfrak{S}_4$. Here we do not go into details of the
construction in  \cite{DS} and just outline some of its key steps.

Suppose the spaces $\V_1$ and $\V_2$ in \p{YB} are infinite-dimensional and the third space
is two-dimensional, $\V_3 = \mathbb{C}^2$. Similar to Sect. \ref{Sklyanin}, two out of three
R-matrices in \p{YB} reduce to the Lax operators
$\mathbb{R}_{13}(u)\equiv\mathrm{L}_{13}(u)$ and
$\mathbb{R}_{23}(u)\equiv\mathrm{L}_{23}(u)$,
but this time they share a common auxiliary two-dimensional space $\V_3= \mathbb{C}^2$.
Matrix entries of $\mathrm{L}_{13}$ are operators acting in $\V_1$ and entries of
$\mathrm{L}_{23}$ act in $\V_2$. Further, we omit the index $3$ in $\mathrm{L}_{i3}, i=1,2,$
of the common auxiliary space.
Because there are two different L-operators
associated with the elliptic modular double, $\mathrm{L}^{doub}$ \p{Lax1} and
$\widetilde{\mathrm{L}}^{doub}$ \p{Lax2},
equation \p{YB} yields two different relations, known as
$\mathrm{RLL}$-relations~\cite{KRS81,skl1}:
\begin{align}\label{RLL0}
\mathbb{R}_{12}(u-v)\,\mathrm{L}^{doub}_1(u)\,\mathrm{L}^{doub}_2(v)=
\mathrm{L}^{doub}_2(v)\,\mathrm{L}^{doub}_1(u)\,\mathbb{R}_{12}(u-v)\,,\\
\mathbb{R}_{12}(u-v)\,\widetilde{\mathrm{L}}^{doub}_1(u)\,\widetilde{\mathrm{L}}^{doub}_2(v)=
\widetilde{\mathrm{L}}^{doub}_2(v)\,\widetilde{\mathrm{L}}^{doub}_1(u)\,\mathbb{R}_{12}(u-v)\,.
\label{RLL01}
\end{align}
The operator $\mathbb{R}_{12}(u) : \V_1 \otimes\V_2 \to \V_1 \otimes\V_2$ is
called the {\it general} R-operator for the elliptic modular double.
Let us stress that it has to respect both operator equations \p{RLL0} and \p{RLL01}
simultaneously. Such an R-operator has been constructed in \cite{DS} by a direct
solution of \p{RLL0} and \p{RLL01} which jointly fix it uniquely up to a normalization constant.

Some more comments are in order. We assume that $\V_1$ is a space of
functions of a complex variable $z_1$ and $\V_2$ is a
space of functions of a complex variable $z_2$. Therefore
the operator $\mathbb{R}_{12}$ acts in the space of functions of two
independent complex variables $z_1$ and $z_2$, $\Phi(z_1,z_2)$,
which form the elements of $\V_1\otimes\V_2$.
We use the shorthand notation: the index $k$ of
$\mathrm{L}^{doub}_k$ and $\widetilde{\mathrm{L}}^{doub}_k$
indicates that the elliptic modular double generators $\mathbf{S}^\alpha_{k}$ and $\widetilde{\mathbf{S}}^\alpha_{k}$
forming these matrices are finite-difference operators \p{SklyanMod}
and \p{mod_doub2} with the spins $g_k$, and they act in the space $\V_k$, i.e.
$\mathbf{S}_k^\alpha\,,\widetilde{\mathbf{S}}^\alpha_{k}\,:\V_k\to\V_k\,$.
The operators $\mathbf{S}_1^\alpha$, $\widetilde{\mathbf{S}}^\alpha_{1}$
and $\mathbf{S}_2^\beta$, $\widetilde{\mathbf{S}}^\beta_{2}$ act
in different spaces and, obviously, they commute with each other,
$\bigl[\mathbf{S}_1^\alpha, \mathbf{S}_2^\beta\bigr] =
\bigl[\widetilde{\mathbf{S}}_1^\alpha, \widetilde{\mathbf{S}}_2^\beta\bigr]
= \bigl[\mathbf{S}_1^\alpha, \widetilde{\mathbf{S}}_2^\beta\bigr]
= \bigl[\widetilde{\mathbf{S}}_1^\alpha, \mathbf{S}_2^\beta\bigr] = 0$.
Matrices $\mathrm{L}^{doub}_k$ in equation~(\ref{RLL0}) are
multiplied as usual $2\times 2$ matrices acting in the space $\V_3=\C^2$,
and the same is valid for $\widetilde{\mathrm{L}}^{doub}_k$ in~(\ref{RLL01}).
In relations \p{RLL0}, \p{RLL01} we omit dependence on the spins $g_k$ in
the L-operators and in the general $\mathrm{R}$-operator.

Let us concentrate our attention to the first RLL-relation \p{RLL0}, i.e.
to a half of the elliptic modular double.
The L-operator for the Sklyanin algebra is not unique.
The transformations $\mathrm{L}^{doub} \to \sigma_{\alpha}\,\mathrm{L}^{doub}$, where $\sigma_\alpha$ is any Pauli matrix,
are automorphisms of the Sklyanin algebra~\cite{skl1}, i.e. they are consistent with the
algebra commutation relations \p{SklAlg}.
Consequently, there are several possible forms of the equation \eqref{RLL0}
with different general $\mathrm{R}$-operators labeled by a supplementary index
$\alpha$ specifying a choice of the L-operators
\be \lb{RLLs3}
\mathbb{R}^{(\alpha)}_{12}(u-v)\,\sigma_\alpha\,\mathrm{L}^{doub}_1(u)
\,\sigma_\alpha\,\mathrm{L}^{doub}_2(v)=
\sigma_\alpha\,\mathrm{L}^{doub}_2(v)\,\sigma_\alpha
\,\mathrm{L}^{doub}_1(u)\,\mathbb{R}^{(\alpha)}_{12}(u-v)\,.
\ee
For a technical reason (see \cite{DS}) we fix $\alpha=3$ from the very beginning and denote
$\mathbb{R}_{jk}(u)\equiv\mathbb{R}^{(3)}_{jk}(u)$ that corresponds to the
$c$-series of the Sklyanin algebra representations \cite{skl2}.
In this case it is possible to cancel one of the $\sigma_3$-matrices such that
the defining $\mathrm{RLL}$-relation takes the form
\begin{equation}\label{RLL}
\mathbb{R}_{12}(u-v)\,\mathrm{L}^{doub}_1(u)
\,\sigma_3\,\mathrm{L}^{doub}_2(v)=
\mathrm{L}^{doub}_2(v)\,\sigma_3
\,\mathrm{L}^{doub}_1(u)\,\mathbb{R}_{12}(u-v)\,.
\end{equation}

One can start from the L-operators obtained by the transformation $\mathrm{L}^{doub} \to
\mathrm{L}^{doub}\sigma_{\alpha}$. However, this leads to the same equation
\eqref{RLLs3}, because we can cancel $\sigma_\alpha$ on the left-hand side and
multiply by the same matrix from the right. The equivalence of these two
options follows from the fact that
$\sigma_\alpha \,\mathrm{L}^{doub}(u_1,u_2) = c_\alpha(u_1,u_2)
\mathrm{L}^{doub}(u_1- \zeta_\alpha,u_2-\zeta_\alpha) \,\sigma_\alpha,$
where $c_\alpha$ are some simple exponential multipliers and
$\zeta_\alpha$ are half-periods 1/2, $\tau/2$, or $1/2+\tau/2$.

It is helpful to extract the permutation operator $\mathbb{P}_{12}$
from the general $\mathrm{R}$-operator
$\mathbb{R}_{12}(u) \equiv \mathbb{P}_{12}\,\mathrm{R}_{12}(u)$ in \p{RLL}, i.e.
we strip off the operator $\mathbb{P}_{12}$ which swaps $z_1$ and $z_2$,
$\mathbb{P}_{12}\,\Phi(z_1,z_2)=\Phi(z_2,z_1)\,\mathbb{P}_{12}$.
Then the defining equation for the general $\mathrm{R}$-operator takes the following form
\begin{equation}\label{RLL'}
\mathrm{R}_{12}(u-v)\,\mathrm{L}^{doub}_1(u_1,u_2)\,\sigma_3\,\mathrm{L}^{doub}_2(v_1,v_2)=
\mathrm{L}^{doub}_1(v_1,v_2)\,\sigma_3\,\mathrm{L}^{doub}_2(u_1,u_2)\,\mathrm{R}_{12}(u-v)\,,
\end{equation}
where we restored dependence on the spins $g_1$ and $g_2$ in the Lax operators
by means of the parameters $u_1,u_2,v_1,v_2$ (cf. \p{u12doub})
\be \lb{uv}
u_1 =
\frac{u+g_1}{2}\ \,,\
u_2 = \frac{u-g_1}{2}\ \ ,\ \
v_1 =
\frac{v+g_2}{2}\ \,, \
v_2 = \frac{v-g_2}{2}\ \,.
\ee
In \p{RLLs3} we still omit dependence of the general $\mathrm{R}$-operator on the spins $g_1$, $g_2$.
The full-fledged notation would be $\mathrm{R}_{12}(u_1,u_2|v_1,v_2)$ or $\mathrm{R}_{12}(u-v|g_1,g_2)$.
The R-operator satisfying equation \eqref{RLL'}
is a composite operator given by the following product \cite{DS}:
\begin{equation}\label{R}
\mathrm{R}_{12}(u-v) =
\mathrm{S}(u_1-v_2)\,
\mathrm{M}_2(u_2-v_2)\,\mathrm{M}_1(u_1-v_1)\,
\mathrm{S}(u_2-v_1)\, ,
\end{equation}
where the constituting blocks -- {\it elementary
intertwining} operators -- are the following operators
\begin{equation}
[\,\mathrm{S}(a)\,\Psi\,](z_1,z_2)=
\Gamma\left(\pm z_1\pm z_2 + a + \eta+\textstyle{\frac{\tau}{2}}\right)
\,\Psi(z_1,z_2)\,,
\label{S2fin}
\end{equation}
\begin{equation}
[\,\mathrm{M}_1(b)\,\Psi\,](z_1,z_2)= \kappa
\int_0^1\frac{\Gamma(\pm z_1 \pm x -b)}
{\Gamma(-2b,\pm 2x)}\, \Psi(x,z_2)\,dx\,,
\label{S1fin}
\end{equation}
\begin{equation}
[\,\mathrm{M}_2(c)\,\Psi\,](z_1,z_2)= \kappa
\int_0^1\frac{\Gamma(\pm z_2 \pm x -c)}
{\Gamma(-2c,\pm 2x)}\, \Psi(z_1,x)\,dx\,,
\label{S3fin}
\end{equation}
and the normalization constant is
$\kappa = \frac{1}{2}(q;q)_\infty\,(p;p)_\infty$.
Note that the operators $\mathrm{M}_1$ and $\mathrm{M}_2$ are copies of the
intertwining operator of the equivalent representations $\mathrm{M}$ \p{M}.
They act in the spaces of functions of variables $z_1$ and $z_2$, respectively.
One can straightforwardly recast the factorized form \p{R} of the $\mR$-operator
in an integral operator form whose kernel contains in the numerator a
product of 16 nontrivial elliptic gamma functions.

It was shown in~\cite{DS} that the elementary intertwining operators satisfy
the following defining relations, which justify the chosen terminology,
\begin{gather}\label{RLL1}
\mathrm{M}_1(u_1-u_2)\,\mathrm{L}^{doub}_1(u_1,u_2) =
\mathrm{L}^{doub}_1(u_2,u_1)\,\mathrm{M}_1(u_1-u_2)\,,
\\ \label{RLL3}
\mathrm{M}_2(v_1-v_2)\,\mathrm{L}^{doub}_2(v_1,v_2) =
\mathrm{L}^{doub}_2(v_2,v_1)\,\mathrm{M}_2(v_1-v_2)\,,
\\ \label{RLL2}
\mathrm{S}(u_2-v_1)\,\mathrm{L}^{doub}_1(u_1,u_2)\,\sigma_3\,\mathrm{L}^{doub}_2(v_1,v_2)=
\mathrm{L}^{doub}_1(u_1,v_1)\,\sigma_3\,\mathrm{L}^{doub}_2(u_2,v_2)\,
\mathrm{S}(u_2-v_1)\,.
\end{gather}
Using these relations one can check straightforwardly that
the general R-operator \p{R} does satisfy the RLL-relation \p{RLL}.
The indicated operators \p{S2fin}, \p{S1fin}, \p{S3fin} are not the most general
solutions of defining relations \p{RLL1}, \p{RLL3}, \p{RLL2}.
We are free to multiply the integrands of operators $\mathrm{M}_1$, $\mathrm{M}_2$,
and the operator $\mathrm{S}$ by periodic functions of $z_{1,2}$
with the period $\eta$ (see \cite{DS} for details). Resorting to the second
half of the elliptic modular double we are able to completely fix this residual freedom. Indeed,
the elementary intertwining operators $\mathrm{S}$, $\mathrm{M}_1$, $\mathrm{M}_2$
are invariant with respect to the permutation of moduli $\tau \rightleftarrows 2\eta$.
The same is true for their product \p{R} -- the general R-operator. Consequently
the general R-operator \p{R} satisfies the accompanying RLL-relation as well,
which contains $\widetilde{\mathrm{L}}^{doub}$ and
is responsible for the second half of the elliptic modular double
\begin{equation}
\mathrm{R}_{12}(u-v)\,\widetilde{\mathrm{L}}^{doub}_1(u_1,u_2)\,\sigma_3\,\widetilde{\mathrm{L}}^{doub}_2(v_1,v_2)=
\widetilde{\mathrm{L}}^{doub}_1(v_1,v_2)\,\sigma_3\,\widetilde{\mathrm{L}}^{doub}_2(u_1,u_2)\,\mathrm{R}_{12}(u-v)\,.
\end{equation}
As a result, the ambiguous periodic functions should be additionally periodic with respect
to the shifts of $z_{1,2}$  by $\tau/2$. Joining to this analytical constraints the demand
that we are working in the space of periodic functions with the period 1
(which is equivalent to the requirement of analyticity in the multiplicative
variables $e^{2\pi i z_{1,2}}$), we fix these functions to be constants by the
Jacobi theorem on the absence of nontrivial triply periodic functions
(the variables $1, \eta,\tau/2$ should be incommensurate for its validity).

Similar ambiguity phenomenon in the $\mathrm{R}$-operator construction takes place for the
Lie algebra $s\ell_2$ and the quantum algebra $U_q(s\ell_2)$ which is cured automatically
for the group $\mathrm{SL}(2,\C)$ and the modular double
$U_q(s\ell_2) \otimes U_{\tilde q}(s\ell_2)$ (see for example \cite{CDS1} and
references therein).

Finally, one needs checking that the outlined construction of the general R-operator
\p{R} is self-consistent. For that it is necessary to go one step up
in the complexity hierarchy of YBE \p{YB} and reach the highest level.
More precisely, it is necessary to show that the derived R-operator
solves this equation when all three spaces $\V_1$, $\V_2$, $\V_3$
form infinite-dimensional modules of the elliptic modular double.
Let us keep corresponding spins $g_1$, $g_2$, $g_3$ in the generic position.
Stripping off the permutation operators $\mathbb{P}_{ij}$ from \p{YB}
we find the Yang-Baxter relation for the general $\mathrm{R}$-operator
\be \lb{YBR}
\mathrm{R}_{23}(u-v) \,\mathrm{R}_{12}(u) \,\mathrm{R}_{23}(v) =
\mathrm{R}_{12}(v) \,\mathrm{R}_{23}(u)\, \mathrm{R}_{12}(u-v) \, ,
\ee
which is indeed satisfied by \p{R}.
The proof is straightforward and is based on
the following relations for the elementary intertwining operators
\be \lb{MM}
\mathrm{M}_1(a)\,\mathrm{M}_2(b)=\mathrm{M}_2(b)\,\mathrm{M}_1(a)\; ,\quad
\mathrm{M}_k(a)\,\mathrm{S}(a+b)\,\mathrm{M}_k(b)=
\mathrm{S}(b)\,\mathrm{M}_k(a+b)\,\mathrm{S}(a)\;, \quad k = 1,2\,.
\ee
It should be stressed that other relations for these operators
\be \lb{MM2}
\mathrm{M}_1(a)\,\mathrm{M}_1(-a)= \mathrm{M}_2(a)\,\mathrm{M}_2(-a)
= \mathrm{S}(a)\,\mathrm{S}(-a)= \II
\ee
are not necessary for establishing \p{YBR}.
The first two quadratic relations in \p{MM2} are two copies of \p{inv},
and the third relation in \p{MM2} is a consequence of the reflection identity \p{refl}.
Cubic relations in \p{MM} represent an operator form of the star-triangle
relation
\be \lb{strtr}
\textstyle
\Gamma(\pm x \pm z + \alpha+\eta +\frac{\tau}{2})\,
\mathrm{M}_z(\alpha+\beta)\,
\Gamma(\pm x \pm z+ \beta+\eta +\frac{\tau}{2})=
\mathrm{M}_z(\beta)\,
\Gamma(\pm x \pm z+\alpha+\beta+\eta +\frac{\tau}{2})
\,\mathrm{M}_z(\alpha)\,,
\ee
which coincides with the Bailey lemma associated with
an elliptic Fourier transformation \cite{spi:bailey}
(we remind that the subindex $z$ for the integral operator $\mathrm{M}_z$
denotes both -- the variable which one integrates over and the external variable
remaining after the integration).
The latter lemma is a direct consequence of the
following elliptic beta integral evaluation formula discovered in \cite{spi:umn}
\begin{equation}\label{1step}
\kappa
\int_0^1 \mathrm{d} z\, \frac{\Gamma(\pm z\pm z_1-a)}{\Gamma(-2a,\pm 2z)}
\cdot
\Gamma(\pm z\pm z_2+a+b+\eta+{\textstyle\frac{\tau}{2}})
\cdot
\frac{\Gamma(\pm x\pm z-b)}{\Gamma(-2b,\pm 2x)}\, =
\end{equation}
$$
= \Gamma(\pm z_1\pm z_2+b+\eta+{\textstyle\frac{\tau}{2}})\cdot
\frac{\Gamma(\pm x\pm z_1-a-b)}{\Gamma(-2a-2b,\pm 2x)}\,
\cdot\Gamma(\pm x\pm z_2+a+\eta+{\textstyle\frac{\tau}{2}})\,,
$$
where $\kappa=\frac{1}{2}(q;q)_\infty(p;p)_\infty$ and imaginary parts
of the parameters $-a\pm z_1$, $-a\pm x$, $a+b\pm z_2+\eta+\tau/2$ are positive
(the latter constraints on the parameters can be relaxed by the analytical
continuation).
A functional form of the star-triangle relation \eqref{strtr} has been considered
in \cite{BS}.

\subsection{Reductions of the general R-operator to finite-dimensional representations}

Now we proceed to the main subject of our work
and reduce the general $\mR$-operator \p{R} to finite-dimensional representations in the first space
that yields a series of new YBE solutions. The principle possibility
for implementing this reduction is given by the following
intertwining relation with the general $\mR$-operator \p{R}
\begin{equation}
\label{base}
\mathrm{M}_1(u_1-u_2)\,\mathbb{R}_{12}(u_1,u_2|v_1,v_2) = \mathbb{R}_{12}(u_2,u_1|v_1,v_2)\,
\mathrm{M}_1(u_1-u_2)\,,
\end{equation}
which can be proved using the identities~(\ref{MM}), \p{MM2}.
This relation shows that both, the null-space
of the intertwining operator $\mathrm{M}_1(g_1)$ and the image space of
$\mathrm{M}_1(-g_1)$ (recall \p{uv}), are mapped onto
themselves by the R-operator $\R_{12}$. Therefore, if we find invariant finite-dimensional
subspaces of the latter spaces they will be invariant with respect to the action
of the $\mR$-operator itself.

Let us stress that the reduction problem is quite nontrivial. Indeed, the general $\mathrm{R}$-operator \p{R}
is an integral operator acting in a pair of infinite-dimensional spaces, but the reduced $\mathrm{R}$-operator
(in one the spaces) becomes a matrix whose entries are finite-difference operators.
The reduction in both spaces of $\mathrm{R}_{12}$ converts an integral operator into a
plain matrix with numerical coefficients depending on $u-v, g_1, g_2, \tau$, and $\eta$.

In order to single out a finite-dimensional subspace out of the infinite-dimensional
representation in the first space we consider the generating function
$\Gamma(\pm  z_1 \pm  z_3 + u_1-u_2)$ (recall \p{genfun1} and \p{genfun2}),
where $z_3$ is an auxiliary parameter, and act upon it by the $\mR$-operator.
We break down the calculation to several steps according to the factorized form
of the general $\mR$-operator \p{R}.
In the end we choose the spin parameter of the first space to sit on
the two-dimensional discrete lattice (recall \p{uv})
\be \lb{g1int}
g_1 = u_1-u_2 = g_{n,m} = (n+1) \,\eta + (m+1) \, {\textstyle \frac{\tau}{2}} \ \ \ \
\text{with} \ \ \ \ n, m \in \mathbb{Z}_{\geq 0}\, ,
\ee
such that the generating function produces a basis of the finite-dimensional
representation in the first space \p{genfunellip}.
However, for a while we assume $g_1 = u_1-u_2$ to be generic.

We act by the first two factors $\mathrm{M}_1(u_1-v_1)\,\mathrm{S}(u_2-v_1)$
of the $\mathrm{R}$-operator \p{R}
on the function $\Gamma(\pm  z_1 \pm  z_3 + u_1-u_2)\,\Phi(z_2)$,
where $\Phi(z_2)$ is an arbitrary holomorphic test function,
\begin{align}
&\mathrm{M}_1(u_1-v_1)\,\mathrm{S}(u_2-v_1) \cdot
\Gamma(\pm  z_1 \pm  z_3 + u_1-u_2) \,\Phi(z_2) = \frac{\Gamma(2 u_1 - 2 u_2)}{\Gamma(2 v_1 - 2 u_2)} \cdot\\ \notag
&\cdot\,
\mathrm{S}(u_2-u_1)\,\Gamma(\pm  z_1 \pm  z_3 + v_1 - u_2)\,
\Gamma(\pm  z_2 \pm  z_3 + u_1 - v_1 + \eta + {\textstyle \frac{\tau}{2}})\, \Phi(z_2)\,.
 \label{step1}\end{align}
The right-hand side expression was obtained using the elliptic beta integral
evaluation formula~(\ref{1step}).
In order to apply the third factor $\mathrm{M}_2(u_2-v_2)$ \p{S1fin} of the
$\mR$-operator \p{R}, we resort to the relation
\begin{multline*}
\mathrm{M}_2(u_2-v_2)\,\Gamma(\pm  z_1 \pm  z_2 + u_2-u_1 + \eta + \textstyle{\frac{\tau}{2}})\,
\Gamma(\pm  z_2 \pm  z_3 + u_1-v_1 + \eta + \textstyle{\frac{\tau}{2}})\,\Phi(z_2) = \\
= \frac{\Gamma(2 u_2 -2 u_1 + 2\eta + \tau)}{\Gamma(2 v_2 - 2 u_2)}\cdot
\mathrm{M}_1(u_1-u_2-\eta-{\textstyle\frac{\tau}{2}})\,
\Gamma(\pm  z_1 \pm  z_2 + v_2-u_2)\,
\Gamma(\pm  z_1 \pm  z_3 + u_1-v_1 + \eta + \textstyle{\frac{\tau}{2}})\,\Phi(z_1)\,,
\end{multline*}
which follows directly from the integral representation \p{M}
for $\mathrm{M}(g)$.
A merit of the derived formula is that we traded a complicated integral operator
$\mathrm{M}_2(u_2-v_2)$ for $\mathrm{M}_1(u_1-u_2-\eta-{\textstyle\frac{\tau}{2}})$,
which turns to the finite-difference operator $\mathrm{M}_1(n\eta+m\frac{\tau}{2})$
\p{genform} for the spin $g_1$ values lying on the lattice \p{g1int}.
Incorporating into the latter formula
the inert factors from \p{step1} and the
last factor $\mathrm{S}(u_1-v_2)$ (see \p{S2fin})
of the $\mR$-operator \p{R}, we find
\begin{eqnarray*} &&
\mathrm{R}_{12}(u_1,u_2|v_1,v_2)\,\Gamma(\pm  z_1 \pm  z_3 + u_1-u_2) \,\Phi(z_2)=
\\  && \makebox[2em]{}
=\frac{1}{\Gamma(2 v_1 - 2 u_2)} \cdot
\mathrm{S}(u_1-v_2)\,\Gamma(\pm  z_1 \pm  z_3 + v_1 - u_2)
\frac{1}{\Gamma(2 v_2 - 2 u_2)}\cdot
\\  && \makebox[4em]{} \cdot
\mathrm{M}_1(u_1-u_2-\eta-{\textstyle\frac{\tau}{2}})\,
\Gamma(\pm  z_1 \pm  z_2 + v_2-u_2)\,
\Gamma(\pm  z_1 \pm  z_3 + u_1-v_1 + \eta +
\textstyle{\frac{\tau}{2}})\,\Phi(z_1)\,.
\end{eqnarray*}
Finally, we apply the reflection relation
$\Gamma(\pm  x+\eta+{\textstyle\frac{\tau}{2}}) = 1$
for the elliptic gamma function \p{refl}, go back to the spin variables \p{uv},
supplement $\mathrm{R}_{12}$ with the permutation operator $\mathbb{P}_{12}$, and
put the spin $g_1$ onto the two-dimensional discrete lattice \p{g1int}:
$g_1 = g_{n,m}= (n+1) \,\eta + (m+1) \, {\textstyle \frac{\tau}{2}}$, at $n, m \in \mathbb{Z}_{\geq 0}$.
Then the previous formula takes the form (after setting $v=0$)
\begin{eqnarray}\nonumber  &&
\mathbb{R}_{12}(u|g_1 = g_{n,m} \,,\,g_2)\,\Gamma(\pm  z_1
\pm  z_3 + g_{n,m})\, \Phi(z_2) =
c \,\cdot\,
\frac{\Gamma(\pm  z_2 \pm  z_3 -\frac{u}{2}+\frac{g_{n,m}+g_2}{2})}
{\textstyle\Gamma(\pm  z_1\pm  z_2 -\frac{u}{2}-\frac{g_{n,m}+g_2}{2}
+ \eta+\frac{\tau}{2})}\,  \cdot
\\   && \makebox[4em]{}
\cdot \mathrm{M}_2( n \,\eta + m \, {\textstyle\frac{\tau}{2}})\,
\frac{\Gamma(\pm  z_1 \pm  z_2 -\frac{u}{2}+\frac{g_{n,m}-g_2}{2})}
{\Gamma(\pm  z_2 \pm  z_3 -\frac{u}{2}+\frac{g_2-g_{n,m}}{2} + \eta +
\textstyle\frac{\tau}{2})}\,\Phi(z_2)\,,
 \lb{redsl2''}\end{eqnarray}
where the finite-difference operator $\mathrm{M}_2( n \,\eta + m \, {\textstyle\frac{\tau}{2}})$
acting in the second space
is given explicitly in \p{genform} and the normalization constant $c$ is
$$
c^{-1} = \Gamma(-u+g_{n,m} \pm  g_2)\,.
$$

Formula \p{redsl2''} constitutes the central result of this paper. It defines a rich class of
solutions of the YBE \p{YB}, which are endomorphisms on the tensor product of
finite-dimensional and infinite-dimensional representations of the
elliptic modular double belonging to the series
specified in Sect. \ref{SecDoub}.
As mentioned at the beginning, this result is obtained without following the
complications with the contours of integration for $\mathrm{M}(g)$-operators which
should always be chosen in an appropriate way. However, in Sect. 6 we confirm expression
 \p{redsl2''} by the fusion procedure, making it thus completely rigorous.
We stress that the proof that the operator \p{R} satisfies YBE \p{YBR}
for some region of spectral parameter and spin values given in \cite{DS} does
not use the inversion relation for $\mathrm{M}(g)$.
Therefore the constraints for the corresponding integration contours can be relaxed
by keeping parameters substantially away from
the critical discrete values leading to pinching of contours,
which substantially simplifies the analysis.

Some comments how to apply formula \p{redsl2''} are in order. Expressions on
both sides of equality \p{redsl2''} depend on the auxiliary parameter $z_3$.
We expand them with respect to the auxiliary basis $\varphi^{(n,m)}_{j,l}(z_3)$ \p{phi}
that induces an expansion of the left-hand side
with respect to the dual basis $\psi^{(n,m)}_{j,l}(z_1)$ \p{psi} of the first space (see \p{genfunellip}).
The finite-difference operator $\mathrm{M}_2( n \,\eta + m \,
{\textstyle\frac{\tau}{2}})$ \p{genform} acts from the left-hand side
and shifts the arguments in the elliptic gamma functions as well as in
the test function $\Phi(z_2)$ by multiples of $\eta$ and $\tau/2$.
After the shifts are performed we can trade all gamma functions for theta functions
by means of the equations \p{Gshift} and the reflection identity \p{refl}.
Expanding the product of theta functions on the right-hand side of \p{redsl2''}
with respect to the auxiliary basis $\varphi^{(n,m)}_{j,l}(z_3)$, we find explicitly
$$
\mathbb{R}_{12}(u|g_{n,m} \,,\,g_2)\,\psi^{(n,m)}_{j,l}(z_1)\,\Phi(z_2)\,.
$$
Expanding this expression in the basis $\varphi^{(n,m)}_{j,l}(z_1)$ by means of
the identities \p{identities}, we obtain a matrix $\left[\,\mathbb{R}_{12}\,
\right]^{r,s}_{j,l}$ of the operator $\mathbb{R}_{12}$ reduced with respect
to the first space written in a pair of bases $\psi^{(n,m)}_{j,l}(z_1)$ and
$\varphi^{(n,m)}_{j,l}(z_1)$, i.e.
$$
\mathbb{R}_{12}(u|g_{n,m} \,,\,g_2)\,\psi^{(n,m)}_{j,l}(z_1) =
\varphi^{(n,m)}_{r,s}(z_1) \Bigl[\,\mathbb{R}_{12}(u|g_{n,m} \,,\,g_2)\,\Bigr]_{j,l}^{r,s}\, ,
$$
where summation over the repeated indices $r,s$ is assumed.
Let us stress that entries of the derived matrices $[\mathbb{R}_{12}]_{j,l}^{r,s}$
are finite-difference operators acting in the second space.
One can find an invertible matrix $C$ providing
the interbasis expansion
$$
\psi^{(n,m)}_{j,l}(z)=\sum_{k=0}^n\sum_{s=0}^mC_{jl,ks}\varphi^{(n,m)}_{k,s}(z)
$$
and pass from the $\mathbb{R}_{12}$-operator matrix form given above to the one
obtained purely in terms of the single basis $\varphi^{(n,m)}_{j,l}(z_1)$ expansion.

After the reduction to a finite-dimensional representation in the first space is
implemented,
we can easily perform reduction to a finite-dimensional representation in the second space as well.
In order to achieve this goal we just substitute the test function $\Phi(z_2)$ for the
generating function of finite-dimensional
representations in the second space \p{genfun1}. Indeed, we choose spins in the first space as
$g_1 = g_{n,m}= (n+1) \,\eta + (m+1) \, {\textstyle \frac{\tau}{2}}$ and in the second space as
$g_2=g_{k,l}= (k+1) \,\eta + (l+1) \, {\textstyle \frac{\tau}{2}}$,
for $n, m ,k ,l \in \mathbb{Z}_{\geq 0}$. Then, the formula \p{redsl2''} yields
\begin{gather}
\mathbb{R}_{12}(u|g_{n,m} \,,\,g_{k,l})\,\Gamma(\pm  z_1 \pm  z_3 + g_{n,m})\, \Gamma(\pm  z_2 \pm  z_4 + g_{k,l})
= c \,\cdot
\frac{\Gamma(\pm  z_2 \pm  z_3 -\frac{u}{2}+\frac{g_{n,m}+g_{k,l}}{2})}
{\Gamma(\pm  z_1\pm  z_2 -\frac{u}{2}-\frac{g_{n,m}+g_{k,l}}{2} + \eta+\textstyle{\frac{\tau}{2}})}\,
\cdot \notag \\ \cdot \,
\mathrm{M}_2( n \,\eta + m \, {\textstyle\frac{\tau}{2}})\,
\frac{\Gamma(\pm  z_1 \pm  z_2-\frac{u}{2}+\frac{g_{n,m}-g_{k,l}}{2})}
{\Gamma(\pm  z_2 \pm  z_3 -\frac{u}{2}+\frac{g_{k,l}-g_{n,m}}{2} + \eta +
\textstyle{\frac{\tau}{2}})}\,\Gamma(\pm  z_2 \pm  z_4 + g_{k,l})\,.\lb{redsl2}
\end{gather}
Here $z_3$ and $z_4$ are auxiliary parameters.
Applying this formula one can straightforwardly extract the finite-dimensional
$\mathbb{R}$-matrices with respect to both spaces in a pair of bases
$\psi^{(n,m)}_{i_1,j_1}(z_1)\,\psi^{(k,l)}_{i_2,j_2}(z_2)$ and
$\varphi^{(n,m)}_{r_1,s_1}(z_1)\,\varphi^{(k,l)}_{r_2,s_2}(z_2)$, i.e.
$$
\mathbb{R}_{12}(u|g_{n,m} \,,\,g_{k,l})\,\psi^{(n,m)}_{i_1,j_1}(z_1)\,\psi^{(k,l)}_{i_2,j_2}(z_2) =
\varphi^{(n,m)}_{r_1,s_1}(z_1) \,\varphi^{(k,l)}_{r_2,s_2}(z_2)
\Bigl[\,\mathbb{R}_{12}(u|g_{n,m} \,,\,g_{k,l})\,\Bigr]_{i_1,j_1,i_2,j_2}^{r_1,s_1,r_2,s_2}\,.
$$
This relation defines new solutions of YBE given as plain
square matrices with  $n m k l$ rows and columns with the elliptic
function entries.

The derived reduction formulae have direct analogues for rational
solutions of YBE as well as in the case of their trigonometric deformation which
were obtained in \cite{CDS1}. As to the Lie group $\mathrm{SL}(2,\mathbb{C})$,
the analogues of \p{redsl2''} and \p{redsl2} are given by
formulae (2.32) and (2.36) in \cite{CDS1},
respectively. In the case of the modular double $U_q(s\ell_2)\otimes U_{\tilde q}(s\ell_2)$
the analogues of \p{redsl2''} and \p{redsl2} are given by formulae (3.36) and (3.37) in
\cite{CDS1}, respectively.

In Sect. \ref{SecFD} we considered two discrete lattices of $g$-spin corresponding to
finite-dimensional representations \p{lattice}. Reduction of the general $\mR$-operator
to the second lattice
$$
g_1 = g_{n,m} = \frac{1}{2} + (n+1) \eta + (m+1) \frac{\tau}{2} \;\;, \qquad n,m \in \mathbb{Z}_{\geq 0},
$$
is also possible. All previous calculations are valid in this case as well, since we fix the spin $g_1$
at the last step. Since the resulting reduction formulae are almost the same we will only
indicate the needed modifications.
In the reduction formula \p{redsl2''} we just need substituting the
finite-difference operator on the right-hand side for another one,
$\mathrm{M}_2( \frac{1}{2} + n \,\eta + m \, {\textstyle\frac{\tau}{2}})$ (see \p{M1/2}).
The generating function of finite-dimensional representations on the left-hand side
of \p{redsl2''} automatically takes the correct form \p{genfun2}.
The same remark is true for the reduction formula \p{redsl2}.
It was mentioned already above that the bases $\varphi^{(n,m)}_{j,l}$
and $\psi^{(n,m)}_{j,l}$ are valid for both lattices.

As to the boundary spin values $g=n\eta,\, n>0,$ or $g=m\tau/2,\, m>0,$
leading to the intertwiner null-spaces,
we do not consider reductions of the general R-operator in these cases, because they do not
yield finite-dimensional R-matrices.

\subsection{Two-dimensional reduction and the L-operator}
\lb{SectLred}

In order to illustrate explicitly how the reduction formula \p{redsl2''} works,
consider reduction of the general $\mR$-operator to Sklyanin's L-operator \p{Lax1}
going through all details of the calculation.
We pick the spin $g_1 = g_{1,0}= 2 \eta + {\textstyle \frac{\tau}{2}}$ (see \p{g1int}) that corresponds to the two-dimensional
representation in the first space of $\mathbb{R}_{12}$ in \p{redsl2''},
\begin{align}
&\R_{12}(u|g_1 =2\eta + {\textstyle\frac{\tau}{2}}\,,\,g)
\, \Gamma(\pm  z_1 \pm  z_3 + 2\eta +
{\textstyle\frac{\tau}{2}})\,\Phi(z_2)
= \notag \\  &=
c \cdot \frac{\Gamma(\pm  z_2 \pm  z_3-\frac{u-g}{2}+\eta+\frac{\tau}{4})}
{\Gamma(\pm  z_1 \pm  z_2-\frac{u+g}{2}   +   \frac{\tau}{4})}
\, \mathrm{M}_{2}(\eta) \,
\frac{\Gamma(\pm  z_1 \pm  z_2-\frac{u+g}{2}+\eta+\frac{\tau}{4})}
{\Gamma(\pm  z_2 \pm  z_3-\frac{u-g}{2}+\frac{\tau}{4})} \, \Phi(z_2)\,.
\lb{ellipRexpl}
\end{align}
The generating function \p{genfun1} of the two-dimensional representation
in the first space appearing on the left-hand side of this formula,
$$
\Gamma(\pm  z_1 \pm  z_3 + 2\eta +
{\textstyle\frac{\tau}{2}}) = -\frac{1}{2}\mathrm{R}^2(\tau)\,
e^{\pi i\frac{\tau}{2}}
\left(\bar\theta_4(z_1)\,\bar\theta_3(z_3) +
\bar\theta_3(z_1)\,\bar\theta_4(z_3) \right)\,,
$$
yields the basis $\varphi^{(1,0)}(z_1)=\{\bar\theta_4(z_1) \,,\, \bar\theta_3(z_1)\}$ (see \p{phi})
which coincides with the second basis $\psi^{(1,0)}(z_1)$ (see \p{psi}). Thus in this particular situation
we do not need to deal with two different bases.
Consequently, the reduction formula \p{redsl2''} produces automatically a matrix of the $\mR$-operator written
in the single basis $\varphi^{(1,0)}(z_1)$, not a pair of bases.

The next ingredient of \p{redsl2''} --- the finite-difference operator $\mathrm{M}_{2}(\eta)$ --- has the form (see \p{Meta})
$$
\mathrm{M}_{2}(\eta) = e^{\pi \textup{i} z_2^2 /\eta}\,\frac{c_A}{\theta_1(2z_2)}
\left[e^{\eta \partial_{z_2}} - e^{-\eta \partial_{z_2}}\right]
e^{-\pi \textup{i} z_2^2 / \eta}\,.
$$
Using equations for the elliptic gamma function
\p{Gshift}, we simplify the right-hand side expression in \p{ellipRexpl} and obtain
\begin{align}
&\R_{12}(u|2\eta + {\textstyle\frac{\tau}{2}},g)
\left(\bar\theta_4(z_1)\,\bar\theta_3(z_3) +
\bar\theta_3(z_1)\,\bar\theta_4(z_3) \right)\Phi(z_2) = \notag
\\
&= \lambda\frac{\bar\theta_3(z_3)}{\theta_1(2z_2)}\,
e^{\pi \textup{i} z_2^2/ \eta}
\left[\,A(z_2,z_1)\,e^{-\pi \textup{i} (z_2+\eta)^2/ \eta}\,\Phi(z_2+\eta)
-A(-z_2,z_1)\,e^{-\pi \textup{i} (z_2-\eta)^2/ \eta}\,\Phi(z_2-\eta)\,\right] + \notag
\\
&+\lambda\frac{\bar\theta_4(z_3)}{\theta_1(2z_2)}\,
e^{\pi \textup{i} z_2^2/ \eta}
\left[\,B(z_2,z_1)\,e^{-\pi \textup{i} (z_2+\eta)^2/ \eta}\,\Phi(z_2+\eta)
-B(-z_2,z_1)\,e^{-\pi \textup{i} (z_2-\eta)^2/ \eta}\,\Phi(z_2-\eta)\,\right] \lb{R1/2}\, ,
\end{align}
where the functions $A(z_2,z_1)$ and $B(z_2,z_1)$ are given by products of three
theta functions\footnote{We use the shorthand notation
$\theta_1(x\pm y) \equiv \theta_1(x + y) \,\theta_1(x - y)$, etc to avoid bulky formulae.}
\begin{align}
A(z_2,z_1) &= \textstyle\bar\theta_4(-z_2-\frac{u-g}{2}-\eta+\frac{\tau}{4})
\,\theta_1(z_2 \pm z_1-\frac{u+g}{2}+\frac{\tau}{4}),
\notag \\
B(z_2,z_1) &= -\textstyle\bar\theta_3(-z_2-\frac{u-g}{2}-\eta+\frac{\tau}{4})
\,\theta_1(z_2 \pm z_1-\frac{u+g}{2}+\frac{\tau}{4}) \lb{AB}
\end{align}
and the normalization constant $\lambda$ is
$$
\lambda  = -\mathrm{R}^2(\tau)\,c\,c_A\,e^{-2\pi i(u+\eta)   +   \pi i \frac{\tau}{2}}\,.
$$
Due to linear independence of the functions $\bar\theta_4(z_3)$ and $\bar\theta_3(z_3)$
forming the basis $\varphi^{(1,0)}(z_3)$ the expression \p{R1/2}
can be split to a pair of operator relations
\begin{align*}
\R_{12}(u|2\eta + {\textstyle\frac{\tau}{2}},g)
\,\bar\theta_4(z_1) &=
\frac{\lambda}{\theta_1(2z_2)}
e^{\pi \textup{i} z_2^2/ \eta}
\left[A(z_2,z_1)\,e^{\eta\partial_{z_2}}-
A(-z_2,z_1)\,e^{-\eta\partial_{z_2}}\right]
e^{-\pi \textup{i} z_2^2 / \eta}\,,
\\
\R_{12}(u|2\eta + {\textstyle\frac{\tau}{2}},g)
\,\bar\theta_3(z_1) &=
\frac{\lambda}{\theta_1(2z_2)}
e^{\pi \textup{i} z_2^2/ \eta}
\left[B(z_2,z_1)\,e^{\eta\partial_{z_2}}-
B(-z_2,z_1)\,e^{-\eta\partial_{z_2}}\right]
e^{-\pi \textup{i} z_2^2/ \eta}\, ,
\end{align*}
where we imply that both side expressions are applied to a test function $\Phi(z_2)$.

Next we expand the product of $\theta_1$-functions in \p{AB} over the basis functions
$\bar\theta_4(z_1)$ and $\bar\theta_3(z_1)$ by means of the identities \p{identities}
$$\textstyle
2\theta_1(z_2 \pm z_1-\frac{u+g}{2}+\frac{\tau}{4}) =
\bar\theta_3(z_1)\,\bar\theta_4(z_2-\frac{u+g}{2}+\frac{\tau}{4})-
\bar\theta_4(z_1)\,\bar\theta_3(z_2-\frac{u+g}{2}+\frac{\tau}{4}).
$$
That yields an expansion of the functions
$A(z_2,z_1)$ and $B(z_2,z_1)$ \p{AB} in this basis
$$
A(z_2,z_1) = a(z_2)\,\bar\theta_4(z_1)+c(z_2)\,\bar\theta_3(z_1)
\ \ , \
B(z_2,z_1) = b(z_2)\,\bar\theta_4(z_1)+d(z_2)\,\bar\theta_3(z_1)\,.
$$
The coefficients in this expansion are given by the following functions of $z_2$,
\begin{align*}
a(z_2) &\textstyle = -\frac{1}{2}\bar\theta_4(-z_2-\frac{u-g}{2}-\eta+\frac{\tau}{4})\,
\bar\theta_3(z_2-\frac{u+g}{2}+\frac{\tau}{4})\,,
\\
b(z_2) &\textstyle = \frac{1}{2}\bar\theta_3(-z_2-\frac{u-g}{2}-\eta+\frac{\tau}{4})\,
\bar\theta_3(z_2-\frac{u+g}{2}+\frac{\tau}{4})\,,
\\
c(z_2) &\textstyle = \frac{1}{2}\bar\theta_4(-z_2-\frac{u-g}{2}-\eta+\frac{\tau}{4})\,
\bar\theta_4(z_2-\frac{u+g}{2}+\frac{\tau}{4})\,,
\\
d(z_2) &\textstyle = -\frac{1}{2}\bar\theta_3(-z_2-\frac{u-g}{2}-\eta+\frac{\tau}{4})\,
\bar\theta_4(z_2-\frac{u+g}{2}+\frac{\tau}{4})\,.
\end{align*}
Now we are able to write the action of the R-operator on the basis
$\{ \bar\theta_4(z_1)\,,\,\bar\theta_3(z_1)\}$ in a matrix form
with the operator entries acting in the second space
$$
\R_{12}(u|2\eta + {\textstyle\frac{\tau}{2}},g)
\,\left(\bar\theta_4(z_1)\,,\bar\theta_3(z_1)\right) = \lambda\cdot\left(\bar\theta_4(z_1)\,,\bar\theta_3(z_1)\right)\cdot
$$
$$
\cdot\,
e^{\pi \textup{i} z_2^2/ \eta}
\frac{1}{\theta_1(2z_2)}
\left(
\begin{array}{cc}
a(z_2)\,e^{\eta\partial_{z_2}}-
a(-z_2)\,e^{-\eta\partial_{z_2}} & b(z_2)\,e^{\eta\partial_{z_2}}-
b(-z_2)\,e^{-\eta\partial_{z_2}}\\
c(z_2)\,e^{\eta\partial_{z_2}}-
c(-z_2)\,e^{-\eta\partial_{z_2}} & d(z_2)\,e^{\eta\partial_{z_2}}-
d(-z_2)\,e^{-\eta\partial_{z_2}}
\end{array} \right ) e^{-\pi \textup{i} z_2^2/ \eta}\,.
$$
Finally, this formula can be rewritten up to a normalization constant in the
form of Sklyanin's L-operator with the shifted spectral parameter $\mathrm{L}^{doub}(u-{\textstyle \frac{\tau}{2}})\sigma_3$ \p{Lax1},
$$
\R_{12}(u|2\eta + {\textstyle\frac{\tau}{2}},g)
\,\left(\bar\theta_4(z_1)\,,\bar\theta_3(z_1)\right) = -\frac{1}{2}\lambda\cdot\left(\bar\theta_4(z_1)\,,\bar\theta_3(z_1)\right)\cdot
$$
\begin{equation}
\cdot\left(
\begin{array}{cc}
w_0(u-\frac{\tau}{2})\,\mathbf{S}^0+w_3(u-\frac{\tau}{2})\,\mathbf{S}^3 &
w_1(u-\frac{\tau}{2})\,\mathbf{S}^1-\textup{i} w_2(u-\frac{\tau}{2})\,\mathbf{S}^2 \\
w_1(u-\frac{\tau}{2})\,\mathbf{S}^1+\textup{i} w_2(u-\frac{\tau}{2})\,\mathbf{S}^2&
w_0(u-\frac{\tau}{2})\,\mathbf{S}^0-w_3(u-\frac{\tau}{2})\,\mathbf{S}^3
\end{array} \right)\sigma_3,\ \ w_{a}(u) = \frac{\theta_{a+1}(u+\eta)}{\theta_{a+1}(\eta)},
\label{L_op1}\end{equation}
and with generators $\mathbf{S}^a$ in the spin $g$ representation which are specified
in \p{SklyanMod}. It can be checked that
\begin{align} \label{s3}
\mathrm{L}^{doub}(u_1 -\textstyle{\frac{\tau}{2}},u_2 -\textstyle{\frac{\tau}{2}})\,\sigma_3 =
- e^{2\pi\textup{i}(\eta -{\textstyle \frac{\tau}{2}}+ u_1 +u_2) }
\,\sigma_3\,\mathrm{L}^{doub}(u_1,u_2),
\end{align}
i.e. we have the required reduction.

In a similar way the reduction of the $\mathrm{R}$-operator with the spin $g_1 = g_{0,1} = \eta + \tau$
gives rise to the L-operator for the second half of the elliptic modular
double $\widetilde{\mathrm{L}}^{doub}$ \p{Lax2}.

\section{Fusion construction}

Fusion is a generally accepted method of constructing higher-spin finite-dimensional solutions of
YBE out of the fundamental ones. Initially it has been developed for the Lie algebra $s\ell_2$
(as well as for $s\ell_N$) \cite{KRS81} and then generalized to deformed
trigonometric and elliptic rank 1 symmetry algebras \cite{KS81}.

The elliptic higher-spin finite-dimensional
R-matrices found numerous applications in condensed matter models \cite{Jimbo}.
Another motivation for considering fusion in the elliptic case is purely mathematical.
On the one hand, the representation theory of $s\ell_2$ and $U_q(s\ell_2)$ is well
understood. Indeed, in these cases the Cartan subalgebra exists
and Verma modules are formed by raising and lowering generators acting on the lowest
weight states. On the other hand, the representation theory of the Sklyanin algebra
is not elaborated well enough and requires further comprehensive studies.
All four generators of the Sklyanin algebra appear on equal footing \p{SklAlg}, which
is the main obstacle preventing the Verma module construction in the elliptic setting.
The fusion method is tightly related to the structure of representations of the underlying
symmetry algebra \cite{KRS81}. Therefore the fusion construction at the elliptic
level is of special interest since it sheds some light on the
Sklyanin algebra representations.
All steps of the general $\mathrm{R}$-operator construction are essentially
the same for deformed and non-deformed algebras \cite{Derkachov:2007gr,DS}.
In this section we will demonstrate that
curiously enough the fusion recipe for the Sklyanin algebra can
be formulated literally in the same way as for the $s\ell_2$ algebra.

The fusion for integrable models with underlying Sklyanin algebra and its
higher rank generalization specified by the $\mathrm{RLL}$-relation with Belavin's
$\mathrm{R}$-matrix has been extensively studied in a number of papers, e.g.,
\cite{DJMO86,DJKMO87,DJKMO88,HZ90,Tak96}. In \cite{McCFa04} a laborious calculation
implemented explicitly the fusion of two, three, and four Baxter's R-matrices reproducing
the Lax operators \p{L_op} for spin values $1$, $\frac{3}{2}$, and $2$, respectively.
In spite of the progress reached
in these works, the elliptic fusion still looks rather complicated
and not many explicit formulae are available. The aim of this section is to
fill this gap in the rank 1 case by developing a novel approach to fusion
(it is most close to the considerations in \cite{konno2,tak}). Namely,
we construct fusion for the elliptic modular double by generalizing the results
of our previous work \cite{CDS1}.
We explicitly perform fusion of {\it arbitrary} number of Baxter's R-matrices
yielding {\it all} finite-dimensional L-operators \p{L_op} and corresponding {\it all}
finite-dimensional representations of the Sklyanin algebra from the series \p{Sklyan1}.
We resort to symbols of finite-dimensional operators
and the auxiliary spinor notation making our calculation fairly simple.
The computations use the factorization of the symbol of Baxter's R-matrix
which is similar to the L-operator factorization \p{LFact}.

After that we proceed to the fusion of L-operators with infinite-dimensional
representations in the quantum space. In this way we produce higher-spin
$\mathrm{R}$-operators acting on the tensor product of an infinite-dimensional
and arbitrary finite-dimensional representations of the Sklyanin algebra. They
coincide with the solutions given by formula \p{redsl2''} which were obtained by
reductions of the general elliptic $\mathrm{R}$-operator.
In this way we find a complete agreement between two approaches to constructing
finite-dimensional (in one or both tensor space factors) YBE solutions.

\subsection{Fusion of Baxter's R-matrices and finite-dimensional representations
of the Sklyanin algebra} \lb{SecFus}

For the rank 1 symmetry algebras underlying an integrable system
the fusion recipe of~\cite{KRS81,KS81} looks as follows.
One forms an {\it inhomogeneous} monodromy matrix
$\mathrm{T}_{i_1\cdots i_n}^{j_1 \cdots j_n}$
out of $\mathrm{L}$-operators \p{L_op} $\mathrm{L}^{j}_{i}$
(or out of Baxter's R-matrices \p{RBaxter} which are $\mathrm{L}$-operators
for the spin $\frac{1}{2}$ \p{Lred})
multiplying them as operators in the quantum space and
taking tensor products of the auxiliary spaces $\C^2$,
\be \lb{Monodr}
\mathrm{T}_{i_1\cdots i_n}^{j_1 \cdots j_n}(u) =
\mathrm{L}^{j_1}_{i_1}(u) \,\mathrm{L}^{j_2}_{i_2}(u - 2 \eta) \cdots
\mathrm{L}^{j_n}_{i_n}(u - 2 (n-1) \eta) \,,
\ee
and then symmetrizes the monodromy matrix over the spinor indices.
The result $\mathrm{T}_{(i_1\cdots i_n)}^{(j_1 \cdots j_n)}$
is an $\mathrm{R}$-operator which has a higher-spin auxiliary space
and solves the YBE because
the parameters of inhomogeneity are adjusted in a special way.

Constructing higher-spin $\mR$-operators in this way one has to deal
with $\mathrm{Sym} \bigl(\C^2\bigr)^{ \otimes \, n}$,
which is a space of symmetric tensors with a number of spinor indices $\Psi_{(i_1 \cdots i_n)}$.
The usual matrix-like action of the operators $\hat{\mathrm{T}}$ has the form
\be \lb{TPsi}
\left[\hat{\mathrm{T}}\,\Psi\right]_{(i_1\cdots i_n)} =
\mathrm{T}_{(i_1\cdots i_n)}^{(j_1 \cdots j_n)}\, \Psi_{(j_1 \cdots j_n)}\, ,
\ee
where the summation over repeated indices is assumed.
We prefer not to deal with a multitude of spinor indices.
Instead we introduce auxiliary spinors $\lambda=(\lambda_1\,,\lambda_2)$,
$\mu = (\mu_1\,,\mu_2)$ and contract them with the tensors
\be \lb{Tlam}
\lambda_{i_1} \cdots \lambda_{i_n} \,\Psi_{i_1 \cdots i_n} = \Psi(\lambda), \qquad
\lambda_{i_1} \cdots \lambda_{i_n}\,
\mathrm{T}_{i_1\cdots i_n}^{j_1 \cdots j_n}
\,\mu_{j_1} \cdots \mu_{j_n}  = \mathrm{T}(\lambda|\mu)\,.
\ee
Thus the symmetrization over spinor indices is taken into account automatically.
Henceforth, in place of tensors we work with corresponding generating functions
which are homogeneous polynomials of degree $n$ of two variables
\be \lb{PsiHom}
\Psi(\lambda) = \Psi(\lambda_1,\lambda_2), \qquad
\Psi(\alpha\lambda_1,\alpha\lambda_2) = \alpha^n\,\Psi(\lambda_1,\lambda_2)\,.
\ee
$\mathrm{T}(\lambda|\mu)$ is usually called the {\it symbol} of the operator.
In this way the formula \p{TPsi} acquires a rather concise form
\be \lb{TPsispinor}
\left[\hat{\mathrm{T}}\,\Psi\right](\lambda) =
\frac{1}{n!}\,\left.\mathrm{T}(\lambda|\dd_{\mu})
\,\Psi(\mu)\right|_{\mu = 0}\,.
\ee
Note that we do not need in fact to take $\mu = 0$ in \p{TPsispinor}.
The variable $\mu$ disappears automatically on the right-hand side of \p{TPsispinor}
since $\mathrm{T}(\lambda|\mu)$ and $\Psi(\mu)$
have the same homogeneity degree.

There is a dual description of the finite-dimensional matrix-like
action on $\mathrm{Sym} \bigl(\C^2\bigr)^{ \otimes \, n}$ which enables one to
get rid of spinor indices as well.
Let us take a spinor $\lambda$ and form a rank $n$ symmetric tensor
$\Psi_{(j_1\cdots j_n)} = \lambda_{j_1} \cdots \lambda_{j_n}$. Then we interpret
$\mathrm{T}_{(i_1\cdots i_n)}^{(j_1 \cdots j_n)}$ as a matrix of the operator $\hat{\mathrm{T}}$
acting on it in a standard way
$$
\hat{\mathrm{T}}\cdot \lambda_{j_1} \cdots \lambda_{j_n}
= \lambda_{i_1} \cdots \lambda_{i_n}
\mathrm{T}_{(i_1\cdots i_n)}^{(j_1 \cdots j_n)}\,.
$$
Contraction of the latter relation with a symmetric tensor formed by another auxiliary spinor
$\mu_{j_1} \cdots \mu_{j_n}$
yields the relation between two formalisms
\be \lb{That}
\hat{\mathrm{T}}\cdot \langle\lambda | \mu \rangle^n = \mathrm{T}(\lambda|\mu),
\ee
where $\langle\lambda | \mu \rangle = \lambda_1\mu_1+\lambda_2\mu_2$.
According to \p{Tlam}, $\langle\lambda | \mu \rangle^n$ is the symbol of the identity operator
$\delta^{j_1}_{(i_1} \cdots \delta^{j_n}_{i_n)}$ in the space of rank $n$ symmetric tensors.
It is also a generating function of the basis vectors since expanding $\langle\lambda | \mu \rangle^n$ in
$\mu$ we recover all of them. Thus relation \p{That} implies that the operator $\hat{\mathrm{T}}$
acting on the generating function produces the symbol of $\hat{\mathrm{T}}$.

We construct the Lax operator \p{L_op} with a finite-dimensional local quantum space
starting from the Baxter R-matrix.
The latter acts on the tensor product of two spin $\frac{1}{2}$ representations
and it is given by expression \p{RBaxter}.
Following the recipe from \cite{KS81} we form the product of Baxter's R-matrices
\be \lb{Rind}
\mathrm{R}_{(i_1\cdots i_n)}^{(j_1 \cdots j_n)}(u) = \mathrm{Sym}\,\mathrm{R}_{i_1}^{j_1}(u)\,\mathrm{R}_{i_2}^{j_2}(u-2\eta)
\,\cdots\,\mathrm{R}_{i_n}^{j_n}(u-2(n-1)\eta)\,,
\ee
where  $\mathrm{Sym}$ implies symmetrization with respect to $(i_1\cdots i_n)$ and
$(j_1 \cdots j_n)\,$ and the indices refer to the first space of the R-matrix \p{RBaxter},
$$
\mathrm{R}_i^j(u) = \sum_{\alpha=0}^3 w_{\alpha} (u)\,
\left(\sigma_\alpha\right)_i^j \sigma_\alpha,\qquad
w_{\alpha}(u) = \frac{\theta_{\alpha+1}
(u+\eta)}{\theta_{\alpha+1}(\eta)}\,.
$$
In such a way we obtain an operator acting in the space of symmetric rank $n$
tensors, i.e. the space of spin $\frac{n}{2}$ representation, and in the
two-dimensional auxiliary space where the Pauli matrices are acting.
According to \cite{KS81} it respects the Yang-Baxter relation. We are going to
find an explicit expression for $\mathrm{R}_{(i_1\cdots i_n)}^{(j_1 \cdots j_n)}(u)$
for all integers $n>0$.

With that end in view we calculate the symbol of \p{Rind} with respect to the quantum space
contracting \p{Rind} with a number of auxiliary spinors $\lambda$ and $\mu$,
\begin{align} \lb{RSklprod} \notag
\mathrm{R}(u|\lambda,\mu) = & \lambda_{i_1} \cdots \lambda_{i_n}\,
\mathrm{R}_{i_1\cdots i_n}^{j_1 \cdots j_n}(u)
\,\mu_{j_1} \cdots \mu_{j_n} = \\ = & \langle\lambda|\mathrm{R}(u)|\mu\rangle
\langle\lambda|\mathrm{R}(u-2\eta)|\mu\rangle\cdots
\langle\lambda|\mathrm{R}(u-2(n-1)\eta)|\mu\rangle\,.
\end{align}
Let us stress that it is still an operator in the auxiliary space, but
for the sake of brevity we refer to it as a symbol of the $\mathrm{R}$-matrix.
The symbol $\mathrm{R}(u|\lambda,\mu)$ factorizes in a product of Baxter's R-matrix symbols
$\langle\lambda|\mathrm{R}(u)|\mu\rangle = \lambda_{i}\,\mathrm{R}_{i}^{j}(u)\,\mu_{j}$.
The homogeneity \p{PsiHom} implies that the auxiliary spinor has one redundant degree
of freedom that can be eliminated by the scaling transformations $\lambda \to \alpha \lambda$.
Further we constrain $\lambda$ as
$\lambda_1 = \bar\theta_4(z)\,, \lambda_2=\bar\theta_3(z)$ with a new parameter $z$,
but we keep components $\mu_1$ and $\mu_2$ independent.
Let us mention that an isomorphism between the representation space of the
Sklyanin algebra $\Theta^{+}_{2n}$ and the space of symmetric tensors
has been established in \cite{Hasegawa97,tak,Tak96}.

Then, taking into account the matrix form of the Sklyanin algebra generators $\mathbf{S}^a$
in spin $\frac{1}{2}$ representation \p{Sred},
we obtain the symbol $\langle\lambda|\mathrm{R}(u)|\mu\rangle$,
$$
\langle\lambda|\mathrm{R}(u)|\mu\rangle
= \left(%
\begin{array}{ll}
  \langle \bar\theta_4\,, \bar\theta_3 |
  \left(w_0(u)\,\sigma_0+w_3(u)\,\sigma_3\right)|\mu\rangle &
\langle\bar\theta_4\,, \bar\theta_3 |
\left(w_1(u)\,\sigma_1-\textup{i} w_2(u)\,\sigma_2\right) |\mu\rangle \\
\langle\bar\theta_4\,, \bar\theta_3 |\left(w_1(u)\,\sigma_1+\textup{i} w_2(u)\,\sigma_2\right)|\mu\rangle &
\langle\bar\theta_4\,, \bar\theta_3 |
\left(w_0(u)\,\sigma_0-w_3(u)\,\sigma_3\right)|\mu\rangle \\
\end{array}%
\right)=
$$
\be \lb{lRm}
= \frac{1}{\theta_1(2\eta)}\,
\left(
\begin{array}{cc}
w_0(u)\,\mathbf{S}^0+w_3(u)\,\mathbf{S}^3 &
w_1(u)\,\mathbf{S}^1-\textup{i} w_2(u)\,\mathbf{S}^2 \\
w_1(u)\,\mathbf{S}^1+\textup{i} w_2(u)\,\mathbf{S}^2&
w_0(u)\,\mathbf{S}^0-w_3(u)\,\mathbf{S}^3
\end{array} \right) \langle\bar\theta_4\,, \bar\theta_3|\mu\rangle\, .
\ee
Let us note that
$\langle \lambda |\mu\rangle = \langle\bar\theta_4\,, \bar\theta_3 |\mu\rangle = \mu_1\bar\theta_4(z)+\mu_2\bar\theta_3(z)$
is a symbol of the identity operator.
Thus we represented $\langle\lambda|\mathrm{R}(u)|\mu\rangle$ as a matrix with elliptic finite-difference operator
entries acting on the symbol of the identity operator.

Then we need to calculate the product \p{RSklprod} of symbols \p{lRm}.
It seems at first sight to be  an extremely complicated task that demands an extensive use
of numerous identities for Jacobi theta-functions.
Fortunately we will be able to avoid difficult calculations due to a peculiar property of the
symbol \p{lRm}.  It can be factorized to a product of three matrices respecting
particular ordering of $z$ and $\dd$,
$$
\langle\lambda|\mathrm{R}(u)|\mu\rangle
= \frac{1}{\theta_1(2\eta)\theta_1(2 z)} \left(
\begin{array}{cc}
\bar{\theta}_3\left(z - \frac{u}{2} - \eta\right) & -\bar{\theta}_3\left(z+\frac{u}{2} + \eta\right) \\
-\bar{\theta}_4\left(z - \frac{u}{2} - \eta\right) & \bar{\theta}_4\left(z+\frac{u}{2} + \eta\right)
\end{array} \right)
\left(
\begin{array}{cc}
\mathrm{e}^{\eta\dd_{z_1}} &0\\
0 & \mathrm{e}^{-\eta\dd_{z_1} }
\end{array} \right )\cdot
$$
\be \lb{Rfact}
\cdot\left(
\begin{array}{cc}
\bar{\theta}_4\left(z+\frac{u}{2}\right) & \bar{\theta}_3\left(z+\frac{u}{2}\right) \\
\bar{\theta}_4\left(z - \frac{u}{2}\right) & \bar{\theta}_3\left(z -\frac{u}{2}\right)
\end{array} \right) \left.
\langle\bar\theta_4(z_1)\,, \bar\theta_3(z_1)|\mu\rangle
\right|_{z_1=z}\,.
\ee
The latter factorization is similar to the factorization of the $\mathrm{L}$-operator \p{LFact}.
However the ordering of shifting operators in \p{Rfact} results in a shift of the spectral
parameter in the rightmost matrix factor as compared to \p{LFact}.

We appreciate the merits of the factorization \p{Rfact} if we calculate the product of two
consecutive symbols from \p{RSklprod}, i.e.
$\langle\lambda|\mathrm{R}(u)|\mu\rangle
\langle\lambda|\mathrm{R}(u-2\eta)|\mu\rangle$.
Indeed, a pair of adjacent lateral matrices cancels in the product owing to the relation
$$
\left(
\begin{array}{cc}
\bar{\theta}_4\left(z+\frac{u}{2}\right) & \bar{\theta}_3\left(z+\frac{u}{2}\right) \\
\bar{\theta}_4\left(z - \frac{u}{2}\right) & \bar{\theta}_3\left(z -\frac{u}{2}\right)
\end{array} \right)
\left(
\begin{array}{cc}
\bar{\theta}_3\left(z - \frac{u}{2}\right) &
-\bar{\theta}_3\left(z+\frac{u}{2}\right) \\
-\bar{\theta}_4\left(z - \frac{u}{2}\right) &
\bar{\theta}_4\left(z+\frac{u}{2}\right)
\end{array} \right) = 2\,\theta_1(2z)\,\theta_1(u)\,
\left(
\begin{array}{cc}
1 & 0\\
0 & 1
\end{array} \right ),
$$
which is equivalent to identities \p{identities}. Thus we obtain
once again a product of three matrices such that all shift operators are gathered in the middle matrix,
\begin{eqnarray*} &&
\langle\lambda|\mathrm{R}(u)|\mu\rangle
\langle\lambda|\mathrm{R}(u-2\eta)|\mu\rangle =
\frac{2\,\theta_1(u)}{\theta^2_1(2\eta)\theta_1(2 z)}
\left(
\begin{array}{cc}
\bar{\theta}_3\left(z - \frac{u}{2} - \eta\right) & -\bar{\theta}_3\left(z+\frac{u}{2} + \eta\right) \\
-\bar{\theta}_4\left(z - \frac{u}{2} - \eta\right) & \bar{\theta}_4\left(z+\frac{u}{2} + \eta\right)
\end{array} \right) \cdot
\\  && \makebox[4em]{}
\cdot\left(
\begin{array}{cc}
\mathrm{e}^{\eta\dd_{1}+\eta\dd_{2}} &0\\
0 & \mathrm{e}^{-\eta\dd_{1}-\eta\dd_{2}}
\end{array} \right )
\left(
\begin{array}{cc}
\bar{\theta}_4\left(z+\frac{u}{2}-\eta\right) & \bar{\theta}_3\left(z+\frac{u}{2}-\eta\right) \\
\bar{\theta}_4\left(z - \frac{u}{2}+\eta\right) & \bar{\theta}_3\left(z -\frac{u}{2}+\eta\right)
\end{array} \right)\cdot
\\ && \makebox[8em]{} \cdot
\left.\langle\bar\theta_4(z_1)\,, \bar\theta_3(z_1)|\mu\rangle
\langle\bar\theta_4(z_2)\,, \bar\theta_3(z_2)|\mu\rangle
\right|_{z_1=z_2=z}\,.
\end{eqnarray*}
Since the product of two symbols has the same structure as the  single symbol \p{Rfact},
the generalization of the previous result to the product of $n$ symbols \p{RSklprod} is evident
\begin{eqnarray} \nonumber &&
\mathrm{R}(u|\lambda,\mu) = \frac{r_n(u)}{\theta_1(2 z)}\cdot
\left(
\begin{array}{cc}
\bar{\theta}_3\left(z - \frac{u}{2} - \eta\right) & -\bar{\theta}_3\left(z+\frac{u}{2} + \eta\right) \\
-\bar{\theta}_4\left(z - \frac{u}{2} - \eta\right) & \bar{\theta}_4\left(z+\frac{u}{2} + \eta\right)
\end{array} \right)
\left(
\begin{array}{cc}
\mathrm{e}^{\eta\dd_{1}+\ldots+\eta\dd_{n}} &0\\
0 & \mathrm{e}^{-\eta\dd_{1}-\ldots-\eta\dd_{n}}
\end{array} \right )\cdot
\\  \nonumber  && \makebox[4em]{}
\cdot\left(
\begin{array}{cc}
\bar{\theta}_4\left(z+\frac{u}{2}-(n-1)\eta\right) & \bar{\theta}_3\left(z+\frac{u}{2}-(n-1)\eta\right) \\
\bar{\theta}_4\left(z - \frac{u}{2}+(n-1)\eta\right) & \bar{\theta}_3\left(z -\frac{u}{2}+(n-1)\eta\right)
\end{array} \right)\cdot
\\ && \makebox[8em]{}  \cdot
\left.\langle\bar\theta_4(z_1)\,, \bar\theta_3(z_1)|\mu\rangle 
\cdots \langle\bar\theta_4(z_n)\,, \bar\theta_3(z_n)|\mu\rangle 
\right|_{z_1=\ldots=z_n=z}\,,
\lb{Rnprod}\end{eqnarray}
where the normalization factor $r_n(u)= 2^{n-1}\,\theta_1(u)\cdots\theta_1(u-2(n-1)\eta) \,\theta^{-n}_1(2\eta)\,$.
Further we multiply three matrices in the latter formula which have mutually commuting entries,
then act by the shift operators from the left by means of the obvious formula
$$
\mathrm{e}^{\pm\eta(\dd_{1}+\ldots+\dd_{n})}
\left. \langle\bar\theta_4(z_1)\,, \bar\theta_3(z_1)|\mu\rangle 
\cdots \langle\bar\theta_4(z_n)\,, \bar\theta_3(z_n)|\mu\rangle 
\right|_{z_1 =\ldots=z_n=z} =
\mathrm{e}^{\pm\eta\dd} \langle\bar\theta_4(z)\,, \bar\theta_3(z)|\mu\rangle^n\,,
$$
and, finally, implement matrix factorization according to \p{LFact}, this time taking into account
non-commutativity of the matrices' entries,
$$
\mathrm{R}(u|\lambda,\mu) =
\frac{r_n(u)}{\theta_1(2 z)}
\left(
\begin{array}{cc}
\bar{\theta}_3\left(z - \frac{u}{2} - \eta\right) & -\bar{\theta}_3\left(z+\frac{u}{2} + \eta\right) \\
-\bar{\theta}_4\left(z - \frac{u}{2} - \eta\right) & \bar{\theta}_4\left(z+\frac{u}{2} + \eta\right)
\end{array} \right)
\left(
\begin{array}{cc}
\mathrm{e}^{\eta\dd} &0\\
0 & \mathrm{e}^{-\eta\dd}
\end{array} \right )\cdot
$$
\be \lb{RFact}
\cdot\left(
\begin{array}{cc}
\bar{\theta}_4\left(z+\frac{u}{2}-n\eta\right) & \bar{\theta}_3\left(z+\frac{u}{2}-n\eta\right) \\
\bar{\theta}_4\left(z - \frac{u}{2}+n\eta\right) & \bar{\theta}_3\left(z -\frac{u}{2}+n\eta\right)
\end{array} \right)
\langle\bar\theta_4(z)\,, \bar\theta_3(z)|\mu\rangle^n  \,.
\ee
In view of the factorized form of the Lax-operator \p{LFact} the previous formula is equivalent to
\be \lb{RFact2}
\mathrm{R}(u|\lambda,\mu) = r_n(u) \; \mathrm{L}\left({\textstyle\frac{u}{2}}+\eta\,,{\textstyle\frac{u}{2}}-n \eta\right) \,
\langle\bar\theta_4(z)\,, \bar\theta_3(z)|\mu\rangle^n\,.
\ee
Once again the two factorized forms of the symbol, with a particular ordering \p{Rnprod},
and without ordering \p{RFact}, are related to each other by a shift of the
 spectral parameter in the rightmost matrix factor.
In order to recover the Lax operator from the symbol \p{RFact2} we
recall that $\langle\bar\theta_4(z)\,, \bar\theta_3(z)|\mu\rangle^n = \langle \lambda|\mu\rangle^n$
and apply the formula \p{That}. Thus the fusion of $n$ Baxter's R-matrices resulted in
the Lax operator $\mathrm{L}(u + (1-n)\eta)$ \p{L_op} with the Sklyanin algebra generators
in the spin $\ell = \frac{n}{2}$ representation.

It is a well known fact that the realization of the algebra \p{SklAlg}
by elliptic finite-difference operators \p{Sklyan1}
found by Sklyanin \cite{skl2} is highly intricate.
Indeed, the straightforward calculation demonstrating that the operators \p{Sklyan1}
do respect the commutation relations \p{SklAlg} is extremely laborious. The explicit
fusion formulae derived in this section provide
an independent check of the algebra commutation relations.
Let us emphasize once more the crucial role played by the factorization formula \p{Rfact} in our
considerations.

\subsection{Fusion for the elliptic modular double}

In the previous section we ``fused" Baxter's R-matrices.
Now we proceed to the fusion of L-operators in the auxiliary space
which have infinite-dimensional representation in the quantum space.
We will concentrate on the elliptic modular double.
The results for the Sklyanin algebra are particular cases of the latter.

We form the inhomogeneous monodromy matrix (recall \p{Monodr}) out of the Lax operators
of two species, $\mathrm{L}^{doub}$ \p{Lax1} and $\widetilde{\mathrm{L}}^{doub}$ \p{Lax2},
and symmetrize it over indices of the auxiliary $\mathbb{C}^2$ spaces
(recall \p{Monodr}),
\begin{align} \lb{RhighInd}
\bigl[\, \mathrm{R}_{\mathrm{fus}}(u) \,\bigr]^{(j_1 \cdots j_{n+m})}_{(i_1\cdots i_{n+m})}
= &\mathrm{Sym}\,\left(\mathrm{L}^{doub}\sigma_3\right)_{i_1}^{j_1}(u)\,
\left(\mathrm{L}^{doub}\sigma_3\right)_{i_2}^{j_2}(u-2\eta)
\,\cdots\,\left(\mathrm{L}^{doub}\sigma_3\right)_{i_n}^{j_n}(u-2(n-1)\eta)\,\cdot \notag\\
&\cdot\,\left(\widetilde{\mathrm{L}}^{doub}\sigma_3\right)_{i_{n+1}}^{j_{n+1}}
(u-2n\eta)
\cdots \left(\widetilde{\mathrm{L}}^{doub}\sigma_3\right)_{i_{n+m}}^{j_{n+m}}
(u-2n\eta-(m-1)\tau)\,.
\end{align}
In the quantum spaces of Lax operators an infinite-dimensional representation with
the spin $g$ is realized.
$\mathrm{R}_{\mathrm{fus}}(u)$ is a higher-spin $\mathrm{R}$-operator
which solves YBE and acts in the tensor product of an infinite-dimensional spin $g$ representation
and finite-dimensional $(n+1)(m+1)$-dimensional spin
$g_{n,m} = (n+1) \eta + (m+1) \frac{\tau}{2}$, $n, m \in \mathbb{Z}_{\geq 0}$, representation.

In the previous section we benefited a lot from auxiliary spinors and the symbol notation that
considerably simplified the calculation. Working with the elliptic modular double we shall
need twice more auxiliary
spinors: $\lambda$, $\tilde{\lambda}$, $\mu$, $\tilde{\mu}$. In view of homogeneity \p{PsiHom}
the spinors contain redundant degrees of freedom. In this calculation we
will retain only independent variables
parametrizing the components of spinors $\lambda,\mu$ by $a$, $b$,
\be \lb{ellipspinor}
\lambda_1 = \lambda_1(a) = \theta_4(a|\textstyle{\frac{\tau}{2}}) \;,\;
\lambda_2 = \lambda_2(a) = \theta_3(a|\textstyle{\frac{\tau}{2}}) \;,\;
\mu_1 = \mu_1(b) = \theta_3(b|\textstyle{\frac{\tau}{2}}) \;,\;
\mu_2 = \mu_2(b) = \theta_4(b|\textstyle{\frac{\tau}{2}})\,.
\ee
The same formulae hold for $\widetilde{\lambda},\,\widetilde{\mu}$ with
the permutation $2 \eta \rightleftarrows \tau$.
The spinors $\lambda$ and $\tilde{\lambda}$ are algebraically independent for generic $a$,
as well as $\mu$ and $\tilde{\mu}$ for generic $b$.
At this point we bear in mind an isomorphism \cite{Tak96,Hasegawa97} between the space of even theta functions
that the Sklyanin algebra generators act upon and the space of symmetric tensors.

In order to find the symbol of $\mathrm{R}_{\mathrm{fus}}$ \p{RhighInd} with respect to the finite-dimensional
space we firstly need to calculate symbols of the L-operators. Similar to \p{Tlam} we contract
Lax operators \p{Lax1} and \p{Lax2} with a pair of auxiliary
spinors in $\mathbb{C}^2$ space yielding
scalar operators\footnote{We are grateful to D. Karakhanyan
and R. Kirschner for a discussion on this point.}
\be \lb{Lam}
\Lambda(u,\lambda,\mu) = \lambda_{i}\, \left(\mathrm{L}^{doub}\sigma_3\right)^{j}_{i}(u) \, \mu_{j}\;\;,\;\;
\widetilde{\Lambda}(u,\widetilde{\lambda},\widetilde{\mu})
= \widetilde{\lambda}_{i} \,\left(\widetilde{\mathrm{L}}^{doub}\sigma_3\right)^{j}_{i}(u) \, \widetilde{\mu}_{j}\,,
\ee
which are linear combinations of the generators $\mathbf{S}^{a}$ \p{SklyanMod}
and $\tilde{\mathbf{S}}^{a}$ \p{mod_doub2}, respectively, in the spin $g$
representation. The operator $\Lambda$ arose previously in \cite{KrZa97}
in the study of vacuum curves of the elliptic Lax operators. Let us stress
that we consider fusion of the twisted Lax operators, i.e.
$\mathrm{L}^{doub}\sigma_3$ (and $\widetilde{\mathrm{L}}^{doub}\sigma_3$),
that is natural in view of results of the Sect.~\ref{SectLred}.

By means of identities \p{identities} one can gather Jacobi theta functions and rewrite the symbol as follows
\begin{align}
\Lambda(u,\lambda,\mu) =
c\,e^{\pi\textup{i}z^2/\eta}\,\frac{1}{\theta_1(2z|\tau)}\Bigl[
\theta_1(z-u_1\pm a |\tau )
\,\theta_1(z+u_2+\eta+\tau \pm b|\tau)\,
e^{4 \pi i z} e^{\eta \dd} - \notag\\
- \theta_1(-z-u_1\pm a |\tau )
\,\theta_1(-z+u_2+\eta+\tau \pm b|\tau)\,
e^{-4 \pi i z} e^{-\eta \dd} \Bigr]\,
e^{-\pi\textup{i}z^2/\eta}\,\,, \lb{Lam1}
\end{align}
where the normalization constant
$c = -4 e^{2\pi \textup{i} (u - g + 2\eta + \tau)}$. The linear combinations $u_1$, $u_2$
of the spectral parameter $u$ and spin $g$ are defined in \p{u12doub}.
Let us remind the shorthand notation adopted here:
$\theta_1(x\pm y) \equiv \theta_1(x + y)\, \theta_1(x - y)$.
Then we take into account that the intertwining operator of equivalent representations
$\mathrm{M}$ \p{M} simplifies to a finite-difference operator \p{Meta}
on a discrete lattice of spin parameters, in particular
\begin{align} \lb{Weta}
\mathrm{M}_z(\eta) = e^{\pi\textup{i}z^2/\eta}\,\frac{1}{\theta_1(2z|\tau)}
\left( e^{\eta \dd_z} - e^{-\eta \dd_z}\right)
e^{-\pi\textup{i}z^2/\eta}\,\,.
\end{align}
Applying the identity
$$
\Gamma(\pm(z+\eta)\pm a+\lambda) = \mathrm{R}^2(\tau)
e^{2\pi \textup{i} (z+\lambda-\eta)}\,\theta_1(z+\lambda-\eta\pm a|\tau)\,
\Gamma(\pm z \pm a+\lambda-\eta)\, ,
$$
which is a consequence of the recurrence relations \p{Gshift}
and the reflection formula \p{refl},
we rewrite \p{Lam1} in terms of $\mathrm{M}_z(\eta)$ and the elliptic gamma functions
\begin{gather}
\Lambda(u,\lambda,\mu) = c\,\cdot\,
\Gamma(\pm a \pm z+u_1 + 2\eta+\tau) \,\Gamma(\pm b \pm z +\eta - u_2)\,
\mathrm{M}_z(\eta) \,\cdot \notag \\ \cdot\, \Gamma(\pm a \pm z+\eta-u_1)\,
\Gamma(\pm b \pm z + 2\eta+\tau+u_2)\,,\lb{Lam2}
\end{gather}
where the constant $c = -4 \mathrm{R}^{-4}(\tau)\, e^{2\pi \textup{i} (u +\eta)}$.

In fact, the chain of transformations leading from \p{Lam} to \p{Lam2} follows in the counter direction as
compared to the calculation in Sect. \ref{SectLred}. Indeed, there we recovered the Lax operator from
the general $\mathrm{R}$-operator by means of the reduction formula \p{redsl2''}.
The right-hand side of the latter relation
does coincide with \p{Lam2} after appropriate identification of parameters.
The factorized representation \p{Lam2} of the symbol $\Lambda$, where the finite-difference
operator \p{Weta} is sandwiched in between two multiplication by a function operators,
is a shadow of the factorization \p{doub}.
Further we profit from the star-triangle relation \p{strtr}
which we rewrite here in an equivalent form
\be\lb{strtriang}
\textstyle
\mathrm{S}_b(\alpha)\,
\mathrm{M}_z(\alpha+\beta)\,
\mathrm{S}_b(\alpha) =
\mathrm{M}_z(\beta)\,
\mathrm{S}_b(\alpha+\beta)\,
\mathrm{M}_z(\alpha)\, ,
\ee
where the following shorthand notation is adopted
\be \lb{Sb}
\textstyle
\mathrm{S}_b(\alpha) = \Gamma(\pm b \pm z +\alpha+\eta +\frac{\tau}{2})\,.
\ee
Thus, due to \p{strtriang},
the operator \p{Lam2} is equal to
\begin{align} \lb{Lam3}
\Lambda(u,\lambda,\mu) = c\,\cdot\,
\mathrm{S}_a(u_1+\eta+{\textstyle\frac{\tau}{2}})
\, \mathrm{M}_z(u_2+\eta+{\textstyle\frac{\tau}{2}})\,
\mathrm{S}_b(\eta)\,
\mathrm{M}_z(-u_2-{\textstyle\frac{\tau}{2}})
\,\mathrm{S}_a(-u_1-{\textstyle\frac{\tau}{2}})\,.
\end{align}
Obviously, the analogues of equalities \p{Lam2} and \p{Lam3} obtained after the permutation
$2\eta \rightleftarrows \tau$ are true for the second symbol $\widetilde{\Lambda}$ \p{Lam}.
Now we are ready to calculate the symbol of the higher-spin $\mathrm{R}$-operator \p{RhighInd}
which factorizes to a product of L-operators' symbols\footnote{Let us note that
strange looking shifts of the spectral parameter in \p{Lstring}
are owing to an unconventional definition of the Lax operators \p{Lax1} and \p{Lax2}
which include a supplementary shift
of the spectral parameter as compared to the usual definitions.}
\begin{align} \notag
\mathrm{R}_{\text{fus}}(u) = &\Lambda(u)\,\Lambda(u-2\eta)\cdots\Lambda(u-2(n-1)\eta)\,\cdot \\
&\cdot\,\widetilde{\Lambda}(u-2n\eta) \cdots
\widetilde{\Lambda}(u-2n\eta - (m-1) \tau)\,, \lb{Lstring}
\end{align}
where we omit the dependence on auxiliary spinors for brevity.
The second factorized representation of the symbols \p{Lam3} enables us
to calculate straightforwardly the product \p{Lstring}. Indeed,
in the product of two consecutive symbols from \p{Lstring} two adjacent pairs of operator factors cancel out
in view of the reflection formula for the elliptic gamma function \p{refl}.
Thus we obtain an explicit expression for the symbol
\begin{align} \lb{Rsymb}
\mathrm{R}_{\text{fus}}(u) =
\gamma_{nm}\cdot\,
\mathrm{S}_a(u_1+\eta+{\textstyle\frac{\tau}{2}})
\, \mathrm{M}_z(u_2+\eta+{\textstyle\frac{\tau}{2}})\,\,
\mathrm{S}^n_b(\eta)\,\mathrm{S}^m_b({\textstyle\frac{\tau}{2}})
\,\,\cdot \notag\\
\cdot\,
\mathrm{M}_z(-u_2+(n-1)\eta+(m-1){\textstyle\frac{\tau}{2}})\,
\mathrm{S}_a(-u_1+(n-1)\eta+(m-1){\textstyle\frac{\tau}{2}})\, ,
\end{align}
where $\gamma_{nm}$ is a normalization constant,
$$
\gamma_{nm} = (-4)^{n+m} \mathrm{R}^{-4n}(\tau)
\,\mathrm{R}^{-4m}(2\eta)\,e^{\pi \textup{i}(2(n+m)u + 2\eta n(2-n) + \tau m(2-m)-4\eta n m)}\,.
$$
Let us note that multiple cancellations in the product \p{Lstring}
are similar to cancellations which we encountered in calculating the symbol $\mathrm{R}(\lambda,\mu)$
\p{Rnprod} in the previous section. There the factorized form \p{Rfact}
of the building blocks constituting the string \p{RSklprod} played a crucial role as well.

Since we know the symbol \p{Rsymb}, we reconstruct immediately the corresponding operator by
means of the relation \p{TPsispinor}
\be \lb{fusEllDoub}
\left[ \,\mathrm{R}_{\text{fus}}(u) \, \Phi \,\right](\lambda,\widetilde{\lambda}|z)
=  \left. \mathrm{R}_{\text{fus}}(u|\lambda,\widetilde{\lambda},\dd_{\mu},\dd_{\widetilde{\mu}}) \,
\Phi(\mu,\widetilde{\mu}|z) \right|_{\mu = \widetilde{\mu}= 0}\,.
\ee
The test function $\Phi(\lambda,\widetilde{\lambda}|z)$ is homogeneous in spinor
variables $\lambda$, $\tilde{\lambda}$ of degree $n$ and $m$, respectively.
Several comments concerning formula \p{fusEllDoub} are in order. It
contains differentiations with respect to auxiliary spinors that is meaningful since according to
\p{Lstring} the symbol is polynomial in $\mu$, $\tilde{\mu}$. However this property of the symbol
is no more obvious for \p{Rsymb} where a sole parameter $b$ is responsible
for the dependence on both spinors \p{ellipspinor}.
In order to recover polynomiality of \p{Rsymb} we resort to the identities
\begin{align}
\mathrm{S}_b(\eta) =
\Gamma(\pm z \pm b + 2 \eta + {\textstyle\frac{\tau}{2}}) &= - \frac{1}{2}\mathrm{R}^{2}(\tau)\,e^{\frac{\textup{i}\pi}{2} \tau}
\left[ \,\mu_1 \theta_4(z|\textstyle{\frac{\tau}{2}}) + \mu_2 \, \theta_3(z|\textstyle{\frac{\tau}{2}}) \,\right]\,,
\lb{Gmu} \\
\mathrm{S}_b({\textstyle\frac{\tau}{2}}) =
\Gamma(\pm z \pm b + \eta + \tau) &= - \frac{1}{2}\mathrm{R}^{2}(2\eta)\,e^{ \textup{i}\pi \eta}
\left[ \,\widetilde{\mu}_1 \theta_4(z|\eta) + \widetilde{\mu}_2 \, \theta_3(z|\eta) \,\right]\, \lb{Gmu2}
\end{align}
following from  \p{identities}, \p{Gshift}, and \p{refl}.
Thus, expression \p{Rsymb} is homogeneous in $\mu$ and $\tilde{\mu}$ with the homogeneity
degrees $n$ and $m$, respectively.
The polynomiality of \p{Rsymb} on $\lambda$ and $\tilde{\lambda}$ is less obvious
although it is guaranteed by \p{Lstring}.
We will verify it explicitly in a moment.

Moreover, we need to check that the ``fused" higher-spin $\mathrm{R}$-operator \p{fusEllDoub}
is identical with the $\mathrm{R}$-operator that resulted from the reduction formula \p{redsl2''}.
With this in mind we take the generating function of the
finite-dimensional
representations of spin $g_{n,m} = (n+1)\eta + (m+1) \frac{\tau}{2}$,
$n, m \in \mathbb{Z}_{\geq 0}$ (see \p{genfunellip})
and rewrite it in terms of the auxiliary spinors' components (recall \p{ellipspinor})
$$
\lambda_1 = \theta_4(a|\textstyle{\frac{\tau}{2}}) \;,\;
\lambda_2 = \theta_3(a|\textstyle{\frac{\tau}{2}}) \;,\;
\widetilde{\lambda}_1 = \theta_4(a|\eta) \;,\;
\widetilde{\lambda}_2 = \theta_3(a|\eta)\,,
$$
that results in a finite product of linear combinations of theta functions depending on an auxiliary parameter $x$,
\begin{align}
\Gamma(\pm  a & \pm  x + (n+1) \eta + (m+1) \textstyle{\frac{\tau}{2}}) = \notag \\
= c_{nm} \,\cdot &
\sideset{}{_{r= 0}^{n-1}}\prod
\left[ \,\lambda_2 \,\theta_4(x+(n-1-2r)\eta\,\big|{\textstyle\frac{\tau}{2}})
+ (-1)^m \,\lambda_1 \,\theta_3(x+(n-1-2r)\eta\,\big|{\textstyle\frac{\tau}{2}}) \,\right] \cdot \notag
\\ \cdot &
\sideset{}{_{s = 0}^{m-1}}\prod
\left[ \,\widetilde{\lambda}_2 \,\theta_4(x+(m-1-2s){\textstyle\frac{\tau}{2}}\,\big|\eta)
+ (-1)^n \,\widetilde{\lambda}_1 \,\theta_3(x+(m-1-2s){\textstyle\frac{\tau}{2}}\,\big|\eta) \,\right].\lb{GenFunn}
\end{align}
The generating function has the homogeneity degrees $n$ and $m$ in $\lambda$ and $\tilde{\lambda}$, respectively.
Now we act on the generating function by $\mathrm{R}_{\mathrm{fus}}$  in the first space according to \p{fusEllDoub}.
Let us stress that the second space of $\mathrm{R}_{\mathrm{fus}}$ is untouched.
The derivatives with respect to spinors in \p{fusEllDoub} are easily calculated by means of the formula
\begin{align} \lb{difSpin}
\left.\mathrm{S}^n_b(\eta)\,\mathrm{S}^m_b({\textstyle\frac{\tau}{2}})
\right|_{\mu\to\dd_{\mu},\widetilde{\mu}\to\dd_{\widetilde{\mu}}}
\Gamma(\pm x \pm b + (n+1) \eta + (m+1) \textstyle{\frac{\tau}{2}}) = \notag \\ =
n!m! ( -2)^{-n-m} \mathrm{R}^{2n}(\tau) \,\mathrm{R}^{2m}(2\eta) \,e^{\frac{\pi \textup{i} \tau}{2} n+ \pi \textup{i} \eta m}\,
\Gamma(\pm z \pm x + (n+1) \eta + (m+1) \textstyle{\frac{\tau}{2}})\, ,
\end{align}
which follows from \p{Gmu}, \p{Gmu2}, \p{GenFunn}.
Effectively, the previous formula implies the substitution (recall \p{Sb})
$$
\mathrm{S}^n_b(\eta)\,\mathrm{S}^m_b({\textstyle\frac{\tau}{2}}) \to
\Gamma(\pm z \pm x + (n+1) \eta + (m+1) \textstyle{\frac{\tau}{2}}) = \mathrm{S}_x(n \eta + m \textstyle{\frac{\tau}{2}})
$$
in \p{Rsymb} that yields
\begin{eqnarray}  \nonumber &&
\mathrm{R}_{\text{fus}}(u)\, \Gamma(\pm  a \pm  x + (n+1) \eta + (m+1) \textstyle{\frac{\tau}{2}}) = \varepsilon_{nm}\cdot
\mathrm{S}_a(u_1+\eta+{\textstyle\frac{\tau}{2}})
\, \mathrm{M}_z(u_2+\eta+{\textstyle\frac{\tau}{2}})\,\,
\mathrm{S}_x(n \eta + m \textstyle{\frac{\tau}{2}})
\,\,\cdot
\\  \nonumber && \makebox[4em]{}
\cdot\,
\mathrm{M}_z(-u_2+(n-1)\eta+(m-1){\textstyle\frac{\tau}{2}})\,\,
\mathrm{S}_a(-u_1+(n-1)\eta+(m-1){\textstyle\frac{\tau}{2}})
\\  \nonumber && \makebox[2em]{}
=  \varepsilon_{nm}\cdot\mathrm{S}_a(u_1+\eta+{\textstyle\frac{\tau}{2}})
\mathrm{S}_x(-u_2+(n-1)\eta+(m-1){\textstyle\frac{\tau}{2}})
\,\cdot
\\   && \makebox[4em]{}
\cdot\,
\mathrm{M}_z(n \eta + m \textstyle{\frac{\tau}{2}})\,
\mathrm{S}_x(u_2+\eta+{\textstyle\frac{\tau}{2}})\,
\mathrm{S}_a(-u_1+(n-1)\eta+(m-1){\textstyle\frac{\tau}{2}}),
 \lb{RFus}\end{eqnarray}
where the normalization constant $\varepsilon_{nm}$ is equal to
$$
\varepsilon_{nm}= 2^{n+m} n! m! \,
\mathrm{R}^{-2n}(\tau) \,\mathrm{R}^{-2m}(2\eta)
\,e^{\pi \textup{i} ( 2(n+m)u + 2\eta n(2-n) + \tau m(2-m)
- 4\eta n m + \eta m +\frac{\tau}{2}n)}\,.
$$
At the last step in \p{RFus} we rearranged the operator factors by means of
the star-triangle relation \p{strtriang}.
Finally, renaming the variables $a = z_1$, $z = z_2$, $x = z_3$,
recalling definitions of $\mathrm{S}_x(\alpha)$ \p{Sb} and $u_1$, $u_2$ \p{u12doub},
and applying the reflection formula \p{refl},
we transform \p{RFus} to the reduction formula \p{redsl2''}.
The precise relation (up to a bulky normalization constant which can be immediately
found taking into account normalizations in \p{redsl2''} and \p{RFus}) is
$$
\mathrm{R}_{\text{fus}}(u+g_{n,m}-\eta -\textstyle \frac{\tau}{2}) = \mathbb{R}_{12}(u|g_{n,m} \,,\,g),
$$
where both sides are assumed to be restricted to a finite-dimensional
spin $g_{n,m}$ representation in the first space.
Thus we have found a nice agreement between two approaches to constructing
finite-dimensional (in  one of the spaces) solutions of the YBE.

\section{Conclusion}

Our previous paper \cite{CDS1} and the present one are devoted to construction
of finite-dimensional matrix solutions of the YBE with the plain function entries
as well as to building of quantum L-operators given by finite-dimensional matrices
with differential or finite-difference operator entries. We derived concise
formulae for these objects from reductions of general integral operator
solutions of YBE with a rank 1 symmetry algebra as well as from the fusion procedure.
At the rational level this result perhaps yields
all finite-dimensional solutions of YBE. At the $q$-deformed level we see at least
two missing classes of solutions. Namely, we did not consider the notorious root of
unity cases for the basic parameter $q$, which form its own separate representation
theory world.
Also we did not consider the R-operator and corresponding reductions for the
model suggested in \cite{spi:conm}.

At the elliptic level the situation is
even more complicated. Our consideration misses the cases when the
base parameters $p$ and $q$ are commensurable, which yields the situation richer
than the plain root of unity cases $q^N=1$ and/or $p^M=1$.
Moreover, there exists the Felderhof model \cite{Fel}, which should be properly
understood from our point of view. Namely, it is necessary to clarify from which
integral operator it can be derived as a reduction and what is the whole hierarchy
of YBE finite-dimensional solutions associated to it? It may happen that such solutions
are obtained from the general R-operator of \cite{DS} as a result of a
more intricate reduction procedure than we have considered here.
We hope to address these questions in the future.

Integral operator solutions of YBE play an important role in various
quantum field theories. As mentioned in \cite{CDS1} (see the references
given there), the rational level case is important for the investigation of
high-energy behaviour of the quantum chromodynamics. However, it is not
completely clear where our finite-dimensional reductions appear
within such applications. We mention also a different important role
played by the elliptic hypergeometric integrals \cite{spi:umn,spi:essays}
in field theories. Namely, they emerge as superconformal indices of four-dimensional
supersymmetric gauge theories \cite{DO,SV} and encode an enormous
amount of information about them and corresponding electromagnetic
dualities. Again, it is not clear what particular kind of useful information on
these models is hidden in the finite-dimensional representations of
the elliptic modular double \cite{DS,DS2} and R-matrices we have
constructed in this work.

The most evident application of our results for future
investigations consists in the detailed consideration of
solvable models analogous to the 8-vertex model using the
most general elliptic R-matrix we have derived. There
should exist also other new many-body integrable models in
classical and quantum mechanics related to our results.
In particular, it would be interesting to investigate the
classical Poisson algebra systems associated with them.


\section*{Acknowledgement}

The authors are indebted to the referee for constructive remarks
and helpful comments.
This work is supported by the Russian Science Foundation
(project no. 14-11-00598).

\end{document}